\documentclass[11pt,a4paper]{article}
\usepackage{jheppub}

\usepackage{amsfonts}
\usepackage{amsmath}
\usepackage{amssymb}
\usepackage{slashed}

\def\tr{{\rm Tr}}
\def\det{{\rm det}}

\def\sign{{\rm sign\,}}
\def\dim{{\rm dim\,}}

\def\bea{\begin{eqnarray}}
\def\eea{\end{eqnarray}}
\def\nn{\nonumber}

\def\lmatrix{\left(\begin{array}}
\def\rmatrix{\end{array}\right)}

\def\eps{\varepsilon}
\def\msbar{\overline{\rm MS\kern-0.5pt}\kern0.5pt}
\def\W{W(F_4)}
\def\Wc{W(B_4)}

\title{Hadron structure on the 16-cell honeycomb}

\author{Sandor Katz, }
\author{Daniel Nogradi}

\affiliation{Department of Theoretical Physics, Eotvos University \\ Pazmany Peter setany 1/a, Budapest 1117, Hungary}

\emailAdd{katz@bodri.elte.hu}
\emailAdd{nogradi@bodri.elte.hu}

\abstract{In previous work we started a program of studying non-perturbative QCD on the 4-dimensional 16-cell honeycomb
lattice instead of the usually adopted cubic lattice. It was shown that in several observables much 
smaller cut-off effects are observed, undoubtedly a result of the larger 
space-time symmetry of this discretization. In the present work further consequences are worked out,
relevant for the lattice determination of hadronic structure functions. The key ingredient is the
larger set of irreducible representations of the 16-cell honeycomb symmetry group relative to the cubic
lattice, leading to more protection from mixing and thus less power divergences. A second ingredient
is that more irreducible representations are available for a hadron at rest, which also leads to
computational gain in practice. The analysis is given in full detail for both polarized and
unpolarized and both forward and off-forward matrix elements of twist-2 and twist-3 operators. 
It is shown that a larger
set of moments can be determined with continuum-like renormalization and at rest, than on the cubic
lattice.
}

\begin{document}

\maketitle

\section{Introduction and summary}
\label{intro}

In previous work \cite{Katz:2025ycr} the program of studying lattice QCD with dynamical Wilson-quarks on the
16-cell honeycomb was initiated. 
The main promise of the approach, which was confirmed in the quenched
approximation through both gluonic and fermionic observables, is that the larger space-time symmetry of
the 16-cell honeycomb relative to the usual 4-dimensional cubic lattice leads to smaller cut-off effects.
Indeed, the number of elements of the point group of the 16-cell honeycomb is three times larger than
that of the cubic lattice. 

This enlarged symmetry also underlies the current work: the renormalization properties of particular 
hadronic matrix elements (we will focus on the nucleon throughout)
\cite{Martinelli:1987zd,Martinelli:1988rr,Gockeler:1995wg,Gockeler:1996mu,Gockeler:2000ja,LHPC:2002xzk,Gockeler:2005vw}.
A larger symmetry, even if it is a finite group as it is on any lattice, provides
more protection from mixing with both lower dimensional and same dimensional operators. Less mixing,
especially with lower dimensional operators leading to power divergences in $1/a$, is directly translated 
into computational gain in a simulation. Furthermore, a matrix element at non-zero momentum is inherently more noisy
than a matrix element between nucleons at rest. The 16-cell honeycomb not only leads to less mixing, but
also more operators are measurable at rest.

We will focus on the 
nucleon structure functions $F_{1,2}(x,Q^2)$ and $g_{1,2}(x,Q^2)$ and associated parton distribution
functions (PDF), measured in deep inelastic scattering
\cite{EuropeanMuon:1987isl,E143:1998hbs,SpinMuon:1998eqa,HERMES:2006jyl,COMPASS:2015mhb,E155:2002iec,JeffersonLabHallA:2014gzr,SANE:2018pwx,H1:2015ubc}
and analyzed theoretically within the operator product expansion
\cite{Shuryak:1981kj,Shuryak:1981pi,Jaffe:1989jz,Jaffe:1990qh,Jaffe:1996zw,Lampe:1998eu}. Their moments can be written in terms
of Wilson coefficients and matrix elements of particular operators at leading order in $1/Q^2$. The
renormalization pattern of these operators on the 16-cell honeycomb will be the main result of our
work. The reduced matrix elements are conventionally denoted by $v_n, a_n, d_n$. We will show that on the
16-cell honeycomb $v_{2,3,4}$, $a_{0,1,2}$, $d_{2,3}$ all renormalize the same as in the continuum
and can be measured at rest. For comparison, state-of-the-art determinations of these quantities on the cubic
lattice can be found in
\cite{Alexandrou:2017oeh,Harris:2019bih,Alexandrou:2020sml,Bhattacharya:2020cen,Hackett:2023rif,FlavourLatticeAveragingGroupFLAG:2024oxs,Taggi:2026skn,Mondal:2026zok},
earlier work, including chiral and finite volume extrapolations as well as the pion and light nuclei, in
\cite{Detmold:2001jb,Detmold:2002nf,Detmold:2003tm,Detmold:2005pt,Detmold:2020snb},
and for reviews see \cite{Lin:2017snn,Cichy:2018mum,Constantinou:2020hdm}. All of these conclusions follow from the
irreducible representations of the symmetry group of the 16-cell honeycomb lattice and we give a full
accounting of these. 

The matrix elements in question involve only fermionic operators in the flavor non-singlet case. The
discussion is extended also to flavor singlet operators which necessarily involve, through mixing, purely gluonic
operators \cite{Detmold:2017oqb,Winter:2017bfs,Shanahan:2018pib}, which are also categorized with respect to the irreducible representations of the 16-cell
honeycomb symmetry group. The improvement over the cubic lattice is also observed in this sector.

The aforementioned matrix elements were all forward, but the discussion is also extended to off-forward matrix
elements, generalized form factors (gravitational form factors and more), involving total derivative operators 
\cite{Ji:1996ek,Ji:1996nm,Hagler:2003jd,Gockeler:2003jfa,LHPC:2007blg,Hackett:2023rif,Chen:2006gg,Shanahan:2018pib}. Here again the same improvement over the cubic
lattice is clear, for the same representation theoretic reason as in the forward case.

Thus the 16-cell honeycomb discretization allows a straightforward 
non-perturbative lattice determination of nucleon structure which
otherwise would only be available from fits to experimental data 
\cite{deFlorian:2014yva,Nocera:2014gqa,Hou:2019efy,Bailey:2020ooq,NNPDF:2021njg} or would be
computationally much more expensive or even prohibitively expensive 
on the cubic lattice \cite{Cichy:2018mum,Taggi:2026skn}. The lowest few moments in turn
constrain the $x$-dependence of the PDFs themselves \cite{Detmold:2001dv,Detmold:2003rq}.

It should be mentioned that on the cubic lattice an alternative approach exists, with also continuum-like
renormalization, using the gradient flow and expanding in small flow time
\cite{Narayanan:2006rf,Luscher:2010iy,Luscher:2011bx,Suzuki:2013gza,Makino:2014taa,Monahan:2015lha,Hieda:2016lly,Harlander:2018zpi,Shindler:2023xpd,Shindler:2024grr,Francis:2025pgf}. This, however, introduces a
well-known window problem, $a \ll \sqrt{8t} \ll 1 / \Lambda_{QCD}$ is required so that cut-off effects
are small yet the flow time is also small in physical units. Satisfying the window constraints is
difficult in practice and complicates the continuum limit.

The paper is organized as follows. In section \ref{structurefunctions}, mainly to fix notation, we
briefly review structure functions, their moments and their parametrization by reduced matrix elements of
twist-2 and twist-3 operators. In section \ref{lattices} we define the 16-cell honeycomb lattice and
describe its symmetry group. Throughout the paper all discussions start with the cubic lattice, which is
of course very well-known but we include it to highlight the difference relative to the 16-cell
honeycomb throughout. In sections \ref{irrepc} and \ref{irrep16} all the irreducible representations of
the symmetry groups of the cubic and 16-cell honeycomb lattices, respectively, are constructed. 
In section \ref{restframe} the irreducible representations, which can give non-zero matrix elements for a
nucleon at rest, are derived. The following four sections contain the main results on
the reduced matrix elements
$v_n$ (unpolarized twist-2 operators), $a_n$ (polarized twist-2 operators), $d_n$ (polarized twist-3
operators) and four-fermion operators. Finally we end with a set of conclusions in section \ref{conc}.

If the reader is only interested in how to construct explicitly the operators in the right irreducible
representation in each of the three cases, $v_n, a_n, d_n$, then these are given in (\ref{vnexplicit}),
(\ref{anexplicit}) and (\ref{dnexplicit}), respectively, and the reader may jump straight there. 

It should be mentioned that lattice field theory on the same lattice, under the names body centered
hypercubic lattice or $F_4$ lattice, was studied earlier for gauge theories and naive fermions 
\cite{Celmaster:1982ht,Celmaster:1983jq,Celmaster:1983vy,Celmaster:1983hs,Celmaster:1984rf,Celmaster:1985cv,Celmaster:1986eb} 
and for scalar field theory too \cite{Neuberger:1987kt,Bhanot:1990zd,Bhanot:1990ai}. 

The irreducible representations of the finite groups involved and their decompositions were
computed using GAP \cite{GAP4}.

\section{Hadronic structure functions}
\label{structurefunctions}

We will use normalization such that for a nucleon of momentum $p$ and
spin 4-vector $S$ we have, in Minkowski signature, $S^2 = - E^2$ and 
of course $p \cdot S = 0$. For instance, in the rest frame, with a
nucleon polarized in the $z$-direction we have $S=(0,0,E,0)$ and $p=(0,0,0,E)$.
In the continuum $E=m_N$ and on the lattice there are of course corrections.

Primarily to fix conventions, let us define the hadronic tensor \cite{Jaffe:1996zw,Lampe:1998eu}
\bea
W^{\mu\nu}(p,S,q) = \frac{1}{2\pi} \int d^4 x e^{iqx} \langle p, S| J^{\mu}(x) J^\nu(0)  | p, S \rangle
\eea
in terms of the electromagnetic current $J^\mu(x)$. Splitting it into its symmetric and anti-symmetric
parts,
\bea
W^{\mu\nu}(p,S,q) = W^{(\mu\nu)}(p,q) + W^{[\mu\nu]}(p,S,q),
\eea
only the anti-symmetric $W^{[\mu\nu]}(p,S,q)$ depends on the spin. The structure functions are then
defined by,
\bea
W^{(\mu\nu)}(p,q) &=& \left( -g^{\mu\nu} + \frac{q^\mu q^\nu}{q^2}\right) F_1(x,Q^2) + 
\left( p^\mu - \frac{p\cdot q}{q^2} q^\mu \right) \left(  p^\nu - \frac{p\cdot q}{q^2} q^\nu\right) \frac{F_2(x,Q^2)}{p\cdot q} \nn \\
W^{[\mu\nu]}(p,S,q) &=& \frac{i}{p\cdot q} \eps^{\mu\nu\rho\sigma} q_\rho \left[ S_\sigma
g_1(x,Q^2) + \left( S_\sigma - \frac{S\cdot q}{p\cdot q} p_\sigma \right) g_2(x,Q^2) \right],
\eea
with $Q^2 = - q^2 > 0$ and $x = \frac{Q^2}{2p\cdot q}$. At leading order in the coupling and $1/Q^2$ we
have, for even $n$,
\bea
\int_0^1 dx x^{n-2} F_2(x,Q^2) &=& \sum_q e_q^2 v_n^{(q)} + \ldots \nn \\
2\int_0^1 dx x^n g_1(x,Q^2) &=& \frac{1}{2} \sum_q e_q^2 a_n^{(q)} + \ldots \\
2\int_0^1 dx x^n g_2(x,Q^2) &=& \frac{1}{2} \frac{n}{n+1} \sum_q e_q^2 \left( d_n^{(q)} - a_n^{(q)} \right) + \ldots \nn
\eea
where the sums are over quark flavors and $e_q$ is the charge of each and $\ldots$ contains the non-trivial
$O(\alpha_s)$ parts of the Wilson coefficients as well 
as $m_N^2/Q^2$ corrections. In what follows the $q$-index on the
quantities $v_n, a_n, d_n$ will be dropped for simplicity. They are related to QCD matrix elements, up to
kinematical factors, of twist-2 and twist-3 operators. The subtraction in the moments of $g_2$ corresponds to the 
general separation of the twist-3 component, $d_n$, from the twist-2 component, $a_n$, via the
Wandzura-Wilczek decomposition \cite{Wandzura:1977qf},
\bea
g_2(x) = g_2^{WW}(x) + {\bar g}_2(x), \qquad \qquad g_2^{WW}(x) = - g_1(x) + \int_x^1 \frac{dy}{y} g_1(y).
\eea
In sections \ref{twist2sec}, \ref{twist2polsec} and \ref{twist3sec} 
the operators will be given for the reduced matrix elements $v_n, a_n, d_n$, which also have well-defined
meaning for odd $n$.

\section{The cubic and the 16-cell honeycomb lattices}
\label{lattices}

It is useful to discuss both lattices in parallel which will highlight the improved properties of the
16-cell honeycomb in a transparent way. Let us choose the usual $e_\mu$ orthonormal 
basis vectors along the 4 Euclidean axes. The 4-dimensional cubic lattice is the set of points with
arbitrary integer, $n_\mu \in \mathbb Z$, coefficients: $n_\mu e_\mu$. In terms of the cubic lattice,
the 16-cell honeycomb consists of two copies, the original one and another one shifted by 
$(\tfrac12,\tfrac12,\tfrac12,\tfrac12)$. This realization often goes by the name body centered
cubic lattice \cite{Celmaster:1982ht}. Another realization, by the name $D_4$-lattice, imposes $\sum n_\mu = $ even, on the cubic
lattice, which is indeed the same as the body centered cubic lattice scaled by $\sqrt{2}$. For
concreteness, if a specific parametrization is needed, we will use body centered cubic coordinates.

It is worthwhile to point out that both in 2 and 4 dimensions there are exactly three regular space
filling tessellations. In 2 dimensions these are the square, triangular and hexagonal lattices. In 4
dimensions they are the cubic, 16-cell and 24-cell honeycombs. The 16-cell honeycomb is analogous to the
triangular lattice and the 24-cell to the hexagonal one. The latter ones are special in the sense that
from point to point the nearest neighbor directions are not the same. Just as the triangular lattice has
a larger symmetry group than the square, the 16-cell honeycomb lattice has a larger symmetry group than the 
cubic. Since the symmetry groups of the 16-cell and 24-cell honeycomb lattices are identical, 
everything in this paper equally applies to the 24-cell honeycomb as well.

On the cubic lattice there are 8 nearest
neighbors, $\pm e_\mu$ at distance 1 and there are 24 next-to-nearest neighbors, $\pm e_\mu \pm e_\nu$
with $\mu < \nu$ at distance $\sqrt{2}$. On the 16-cell honeycomb there are 
8+16=24 nearest neighbors, $\pm e_\mu, (\pm\tfrac12,\pm\tfrac12,\pm\tfrac12,\pm\tfrac12)$ at distance 1,
and also 24 next-to-nearest neighbors, $\pm e_\mu \pm e_\nu$ with $\mu < \nu$ at distance $\sqrt{2}$.
The nearest neighbors are often called short, the next-to-nearest neighbors long. The larger symmetry is also
manifested by the symmetric arrangement of short and long neighbors on the 16-cell honeycomb, (24, 24), 
relative to the cubic lattice, (8, 24).

More concretely, the symmetry group of each lattice is generated by reflections along both the short and
long neighbors or roots. The full set is called a root system and the $8+24 = 32$ neighbors in the cubic case are
the $B_4$ root system, while the $24+24=48$ neighbors in the 16-cell honeycomb case are the $F_4$ root system. The
full symmetry groups of both lattices, finite subgroups of $O(4)$, are generated by reflections along all neighbors, 
both short and long and is called the Weyl group of the root system, $\Wc$ and $\W$. The notation $H(4) =
\Wc$ is often used, but we will retain $\Wc$ here because of the parallel with $\W$.

In the cubic lattice $\Wc$ can equivalently be described as permutations, $\pi$, of the 4 coordinates as well 
as sign flips of each coordinate by $\eps_\mu = \pm 1$. Hence $|\Wc| = 2^4 \cdot 4! = 384$. Just two elements generate
$\Wc$,
\bea
g_1 = \lmatrix{cccc} 
1 & 0 & 0 & 0 \\
0 & -1 & 0 & 0 \\
0 & 0 & 0 & 1 \\
0 & 0 & 1 & 0 \rmatrix \qquad \qquad \qquad
g_2 = \lmatrix{cccc}
0 & 1 & 0 & 0 \\
0 & 0 & 1 & 0 \\
0 & 0 & 0 & 1 \\
1 & 0 & 0 & 0 \rmatrix,
\label{generators}
\eea
where $g_1^2 = 1$ and $g_2^4 = 1$.

The symmetry group of the 16-cell honeycomb can be described by noting that there are 3 orthonormal sets
of 4 vectors among the 12 positively oriented nearest neighbors. These, grouped in 3, with the 4 vectors
in each group
assembled into a $4\times 4$ matrix as columns, can be chosen as,
\bea
e^{(1)} = &\lmatrix{cccc}
1 & 0 & 0 & 0 \\
0 & 1 & 0 & 0 \\
0 & 0 & 1 & 0 \\
0 & 0 & 0 & 1 \rmatrix& \label{ebasis} \\
e^{(2)} = \frac12 \lmatrix{rrrr}
1 & 1 & 1 & -1 \\
1 & 1 & -1 & 1 \\
1 & -1 & 1 & 1 \\
1 & -1 & -1 & -1 \rmatrix &\qquad& 
e^{(3)} = \frac12 \lmatrix{rrrr}
1 & 1 & 1 & 1 \\
1 & 1 & -1 & -1 \\
1 & -1 & 1 & -1 \\
-1 & 1 & 1 & -1 \rmatrix \nn.
\eea
Now the $SO(4)$ matrix $A = e^{(2)}$ takes one of these bases into the next, $A e^{(1)} = e^{(2)}$, $A
e^{(2)} = e^{(3)}$, $A e^{(3)} = e^{(1)}$. Of course $A^3 = 1$ and so $A$ corresponds to a rotation by
$120^\circ$. This triality structure implies, among other things, that the
number of elements of $\W$ is 3 times larger than $\Wc$ so we have $|\W| = 1152$. Still two generators
are sufficient to generate the entire $\W$, the same $g_1$ as in (\ref{generators}) and
\bea
g_2 = \frac12 \lmatrix{rrrr}
-1 & -1 &  1 &  1 \\
 1 & -1 & -1 &  1 \\
 1 &  1 &  1 &  1 \\
 1 & -1 &  1 & -1 \rmatrix,
\eea
for which we have $g_2^6 = 1$.

The 16-cell honeycomb (and the 24-cell honeycomb) are in fact the most symmetric lattices in 4 dimensions
in the sense that the number of elements of their symmetry group is larger than any other \cite{Conway:1999dov}.

Any finite subgroup of $O(4)$ can be characterized by its invariant polynomials in $p_\mu$. Under $O(4)$
there is a single invariant, $p^2$, of order 2 of course. With respect to $\Wc$ the independent orders are 2, 4, 6,
8 and a basis for the corresponding polynomials, beyond $p^2$, can be chosen as, 
\bea
p^{(2n)} = \sum_\mu p_\mu^{2n}  \qquad \qquad    n = 2,3,4,
\label{polyc}
\eea
while on the 16-cell honeycomb the independent orders are 2, 6, 8, 12 and the following forms a basis,
together with $p^2$,
\bea
I_6 &=& 5 p^2 p^{(4)} - 4 p^{(6)} \nn \\
\label{poly}
I_8 &=& 35 \left(p^{(4)}\right)^2 + 28 p^2 p^{(6)} - 60 p^{(8)} \\
I_{12} &=& 22 p^2 p^{(10)} + 165 p^{(4)} p^{(8)} + 154 \left( p^{(6)} \right)^2 - 340 p^{(12)}. \nn
\eea
For example, the only order 4 invariant polynomial is $(p^2)^2$, which is also $O(4)$ invariant. This
full rotational invariance of the order 4 invariant lattice polynomial was responsible for the lack of
$O(a^2)$ corrections at tree level in various quantities \cite{Katz:2025ycr}.

The decomposition of representations of $O(4)$ under both symmetry groups will be the key ingredient
for all of the physics conclusions. The $O(4)$ representations will be
labeled by $(j_1,j_2) \oplus (j_2,j_1)$ for $j_1 \neq j_2$ and $(j,j)^\pm$. In the latter case $(j,j)^+$
corresponds to the parity invariant $SO(4)$ representation and $(j,j)^- = \det \otimes (j,j)^+$.
The fundamental representation, $R$, is $(\tfrac12,\tfrac12)^+$. The 2-index anti-symmetric
representation will be denoted by $A$ and is equivalent to $(1,0) \oplus (0,1)$.
The space of order $n$ harmonic polynomials, $H_n$, will also play a role. These are equivalent to the
$(\tfrac{n}{2},\tfrac{n}{2})^+$ representation of dimension $(n+1)^2$ and consist of 
polynomials of the form $p_{\mu_1}\cdots p_{\mu_n} - (traces)$. Equivalently, they can be described as
the space of $n$-index totally symmetric traceless tensors. Another representation which will show
up, $M_{n+1}$, corresponding to the Young tableau $[n,1]$, consists of the $(n+1)$-index tensors with mixed
$[n,1]$-type symmetry. The dimension is $2n(n+2)$ and corresponds to
$(\tfrac{n+1}{2},\tfrac{n-1}{2}) \oplus (\tfrac{n-1}{2},\tfrac{n+1}{2})$. Yet another representation
which will be relevant is a 4-index traceless tensor representation, $W$, subject to the symmetries and anti-symmetries
of the curvature tensor of general relativity (Weyl tensor). This is equivalent to $(2,0) \oplus (0,2)$.

The irreducible 2-dimensional representation of the $S_3$ group, permutations of 3 elements, will show up
in the discussion of both symmetry groups, so we will describe it here. On the 3-vector, $(x,y,z)$, there
is an obvious 3-dimensional representation of $S_3$ given by permutation matrices. This is not
irreducible however, the plane $x+y+z=0$ is fixed, hence it decomposes into an irreducible 2-dimensional
representation and the trivial one. This 2-dimensional representation of $S_3$ will be labeled by ${\bf
2}$ with various subscripts for context.

All of the physical consequences will follow from the irreducible representations of $\Wc$ and $\W$ and
how the above listed $O(4)$ representations decompose under these two finite subgroups. The case
of the cubic lattice is of course well-known, but we will include it in the next section for
completeness and in order to contrast it with the 16-cell honeycomb in a self-consistent way.

The twist-2 and twist-3 operators that will play a role in sections \ref{twist2sec}, \ref{twist2polsec}
and \ref{twist3sec} and 
all involve the covariant derivative $D_\mu$. On the cubic lattice, where the positively oriented
four nearest neighbor directions
are along the Cartesian directions, it is discretized straightforwardly. On the 16-cell honeycomb, on the
other hand, we have 12 positively oriented nearest neighbors, $a=1,\ldots,12$, and we have $D_\mu =
\tfrac13 e^a_\mu D_a$ where $e^a$ are the 12 columns of $e^{(1,2,3)}$ from (\ref{ebasis}) and $D_a$ is
the straightforwardly discretized covariant derivative in the $a^{th}$ direction. 

\section{Irreducible representations of the cubic point group}
\label{irrepc}

$\Wc$ has 20 irreducible
representations and we list them below, 
following the notation in \cite{Baake:1981qe}, also adopted in \cite{Gockeler:1996mu}. 
These will be labeled by $\tau^{(k)}_l$ where $k$ is
the dimension of the representation and $l$ labels the different ones for a given $k$. 

First, let us describe each $\tau^{(k)}_l$ in detail. The main building blocks are the 1-dimensional representations,
a particular 2-dimensional one (to be described below), the 3-dimensional traceless diagonal 2-index
representation $D$, the 6-dimensional off-diagonal symmetric 2-index representation $S$, 
and the also 6-dimensional anti-symmetric 2-index representation $A$, mentioned in the previous section
already. 

The 1-dimensional ones are the trivial one, the product of the sign flips $\eps = \prod_\mu \eps_\mu$, 
the sign of the permutation, $\sigma = \sign(\pi)$ and $\sigma \eps = \det$. The 2-dimensional one is 
the following. There are 3 ways to pick a pair of orthogonal planes 
among the basis vectors and $\Wc$ permutes these planes,
leading to a representation of $S_3$. The construction mentioned in the previous section produces an
irreducible 2-dimensional representation, which we call ${\bf 2}$.
From these ingredients the higher dimensional ones are built up and are listed below.
From the lower dimensional ones tensor products produce the higher dimensional ones, taking into
account that some products are identical,
\bea
\sigma \otimes {\bf 2} = {\bf 2}, \qquad
\eps \otimes S = S, \qquad
\det \otimes A = A,
\label{redund}
\eea
so not every product appears. Then the 20 irreducibles are,
\bea
\begin{array}{llll}
    \tau^{(1)}_1 = 1, & \tau^{(1)}_2 = \eps, & \tau^{(1)}_3 = \sigma, & \tau^{(1)}_4 = \det  \\ \\
    \tau^{(2)}_1 = {\bf 2}, &  \tau^{(2)}_2 = \eps \otimes {\bf 2} & &  \\ \\
    \tau^{(3)}_1 = D, & \tau^{(3)}_2 = \eps \otimes D, & \tau^{(3)}_3 = \sigma \otimes D, & \tau^{(3)}_4 = \det \otimes D  \\ \\
    \tau^{(4)}_1 = R, & \tau^{(4)}_2 = \eps \otimes R, & \tau^{(4)}_3 = \sigma \otimes R, & \tau^{(4)}_4 = \det \otimes R  \\ \\
    \tau^{(6)}_1 = A, & \tau^{(6)}_2 = \sigma \otimes A & & \\ \\
    \tau^{(6)}_3 = S, & \tau^{(6)}_4 = \sigma \otimes S & & \\ \\
    \tau^{(8)}_1 = {\bf 2} \otimes R, & \tau^{(8)}_2 = \eps \otimes {\bf 2} \otimes R & & \\ 
\label{taudef}
\end{array}
\eea
Note that $4 \cdot 1 + 2 \cdot 2^2 + 4 \cdot 3^2 + 4 \cdot 4^2 + 4 \cdot 6^2 + 2
\cdot 8^2 = 384 = |\Wc|$, as it should be, following the general sum rule over all irreducible
representations $\rho$,
\bea
\label{sumr}
\sum_\rho (\dim \rho)^2 = |G|\;,
\eea
of a finite group $G$.

Most important is how the $O(4)$ representations decompose under $\Wc$ and these are listed in
table \ref{o4decomp}. A useful cross-check is matching of dimensions, which is seen to hold for all
cases.

\begin{table}
\centering
\small
\renewcommand{\arraystretch}{1.6}
\setlength{\tabcolsep}{4pt}
\begin{tabular}{|c|c|c|c|}
\hline
 &  & $\Wc$ & $\W$ \\
\hline
\hline
$(0,0)^{+}$ & $1$ & $\tau^{(1)}_1$ & $\varrho^{(1)}_1$ \\
\hline
$(0,0)^{-}$ & $\det$ & $\tau^{(1)}_4$ & $\varrho^{(1)}_4$ \\
\hline
$(\tfrac12,\tfrac12)^{+}$ & $R$ & $\tau^{(4)}_1$ & $\varrho^{(4)}_1$ \\
\hline
$(\tfrac12,\tfrac12)^{-}$ & $\det \otimes R$ & $\tau^{(4)}_4$ & $\varrho^{(4)}_4$ \\
\hline
$(1,0)\oplus(0,1)$ & $A$ & $\tau^{(6)}_1$ & $\varrho^{(6)}_1$ \\
\hline
$(1,1)^{+}$ & ${H}_2$ & $\tau^{(3)}_1\oplus\tau^{(6)}_3$ & $\varrho^{(9)}_1$ \\
\hline
$(1,1)^{-}$ & $\det \otimes {H}_2$ & $\tau^{(3)}_4\oplus\tau^{(6)}_4$ & $\varrho^{(9)}_4$ \\
\hline
$(2,0)\oplus(0,2)$ & ${W}$ &
$\tau^{(2)}_1\oplus\tau^{(2)}_2\oplus\tau^{(6)}_2$ &
$\varrho^{(4)}_5\oplus\varrho^{(6)}_2$ \\
\hline
$(\tfrac32,\tfrac12)\oplus(\tfrac12,\tfrac32)$ & $M_3$ &
$\tau^{(8)}_1\oplus\tau^{(8)}_2$ & $\varrho^{(16)}_1$ \\
\hline
$(\tfrac32,\tfrac32)^{+}$ & ${H}_3$ &
$\tau^{(4)}_1\oplus\tau^{(4)}_2\oplus\tau^{(8)}_1$ &
$\varrho^{(8)}_1\oplus\varrho^{(8)}_3$ \\
\hline
$(\tfrac32,\tfrac32)^{-}$ & $\det\otimes{H}_3$ &
$\tau^{(4)}_3\oplus\tau^{(4)}_4\oplus\tau^{(8)}_2$ &
$\varrho^{(8)}_2\oplus\varrho^{(8)}_4$ \\
\hline
$(2,2)^{+}$ & ${H}_4$ &
$\tau^{(1)}_1\oplus\tau^{(1)}_2\oplus\tau^{(2)}_1\oplus\tau^{(3)}_1\oplus\tau^{(6)}_1\oplus\tau^{(6)}_2\oplus\tau^{(6)}_3$ &
$\varrho^{(2)}_1\oplus\varrho^{(2)}_3\oplus\varrho^{(9)}_1\oplus\varrho^{(12)}_1$ \\
\hline
$(2,2)^{-}$ & $\det \otimes {H}_4$ &
$\tau^{(1)}_3\oplus\tau^{(1)}_4\oplus\tau^{(2)}_2\oplus\tau^{(3)}_4\oplus\tau^{(6)}_1\oplus\tau^{(6)}_2\oplus\tau^{(6)}_4$ &
$\varrho^{(2)}_2\oplus\varrho^{(2)}_4\oplus\varrho^{(9)}_4\oplus\varrho^{(12)}_1$ \\
\hline
$(2,1)\oplus(1,2)$ & $M_4$ &
$\tau^{(3)}_2\oplus\tau^{(3)}_3\oplus\tau^{(6)}_1\oplus\tau^{(6)}_2\oplus\tau^{(6)}_3\oplus\tau^{(6)}_4$ &
$\varrho^{(9)}_2\oplus\varrho^{(9)}_3\oplus\varrho^{(12)}_1$ \\
\hline
\end{tabular}
    \caption{Decomposition of $O(4)$ representations under $\Wc$ and $\W$. We list the common
    $(j_1,j_2)$ and $(j,j)^\pm$ labels as well as the realizations used in the text, $\det, R, A, H_n, M_n, W$.
    }
\label{o4decomp}
\end{table}

\section{Irreducible representations of the 16-cell honeycomb point group}
\label{irrep16}

The symmetry group of the 16-cell honeycomb, $\W$, has 25 irreducible representations
which we label by
$\varrho^{(k)}_l$, similarly to the labeling $\tau^{(k)}_l$ for the cubic lattice. So $k$ is the
dimension of the representation and $l$ labels the inequivalent ones for the same dimension.

The building blocks are again the low dimensional representations. There are four 1-dimensional ones,
the trivial one, $\chi_s$, $\chi_l$ and $\det$. Here $\chi_l$ and $\chi_s$ count the number of
short and long reflections (see section \ref{lattices} for their description) with signs,
\bea
\chi_s = (-1)^{number\; of\; short\;\; reflections}, \qquad \qquad \chi_l = (-1)^{number\;\; of\;\; long\;\; reflections}.
\eea
We have $\left.\chi_s\right|_{\Wc} = \eps$ and
$\left.\chi_l\right|_{\Wc} = \sigma$ and $\chi_s \chi_l = \det$ similarly to the identity 
$\sigma\eps = \det$ in the cubic lattice case. 

There are two basic 2-dimensional representations, both following the $S_3$ construction in 
section \ref{lattices}. The 12 positively oriented short roots can be arranged into three sets of four basis
vectors which form an orthonormal basis. The group $\W$ permutes these three sets and 
the representation ${\bf 2}_s$ is the corresponding irreducible representation. The same construction
with the long roots gives ${\bf 2}_l$. Then there is the
already mentioned 6-dimensional anti-symmetric 2-index representation $A$ and 
the 9-dimensional traceless symmetric 2-index
representation $H_2$. From these ingredients we get the 25 irreducible representations,
\bea
\begin{array}{llll}
\varrho^{(1)}_1 = 1, & \varrho^{(1)}_2 = \chi_s, & \varrho^{(1)}_3 = \chi_l, & \varrho^{(1)}_4 = \det \\ \\
    \varrho^{(2)}_1 = {\bf 2}_l, & \varrho^{(2)}_2 = \det \otimes {\bf 2}_l, 
    & \varrho^{(2)}_3 = {\bf 2}_s, & \varrho^{(2)}_4 = \det \otimes {\bf 2}_s \\ \\
\varrho^{(4)}_1 = R, & \varrho^{(4)}_2 = \chi_s\otimes R, & \varrho^{(4)}_3 = \chi_l \otimes R, & \varrho^{(4)}_4 = \det \otimes R \\ \\
    \varrho^{(4)}_5 = {\bf 2}_l \otimes {\bf 2}_s & & & \\ \\
\varrho^{(6)}_1 = A, & \varrho^{(6)}_2 = \chi_l \otimes A & & \\ \\
    \varrho^{(8)}_1 = {\bf 2}_l \otimes R, &  
    \varrho^{(8)}_2 = \det\, \otimes {\bf 2}_l \otimes R, &  
    \varrho^{(8)}_3 =  {\bf 2}_s \otimes R, &  
    \varrho^{(8)}_4 = \det\,\otimes {\bf 2}_s \otimes R  \\ \\
\varrho^{(9)}_1 = H_2, & \varrho^{(9)}_2 = \chi_s\otimes H_2, & \varrho^{(9)}_3 = \chi_l \otimes H_2, & \varrho^{(9)}_4 = \det \otimes H_2 \\ \\
    \varrho^{(12)}_1 = {\bf 2}_l \otimes A & & & \\ \\
    \varrho^{(16)}_1 =  {\bf 2}_l \otimes {\bf 2}_s \otimes R  & & & \\ 
\end{array}
\eea
and of course (\ref{sumr}) is satisfied with
\bea
4 \cdot 1^2 + 4 \cdot 2^2 + 5 \cdot 4^2 + 2 \cdot 6^2 + 4 \cdot 8^2 + 4 \cdot 9^2 + 1 \cdot 12^2 + 1\cdot
16^2 = 1152\;. \nn
\eea
There are again products which are identical and hence are already included in the above table,
\bea
\chi_s \otimes {\bf 2}_s = {\bf 2}_s, &\qquad&  \chi_l \otimes {\bf 2}_l = {\bf 2}_l, \\ \nn \\
\chi_l \otimes A = \chi_s \otimes A, \qquad  &\det \otimes A = A,& \qquad {\bf 2}_s\otimes A = {\bf 2}_l\otimes A. \nn
\eea
The decompositions of the $O(4)$ representations under $\W$ are listed in table \ref{o4decomp}.
The 25 representations further decompose under $\Wc \subset \W$ and these are also listed below,
\bea
\label{rhotau}
\begin{array}{llll}
\varrho^{(1)}_1 = \tau^{(1)}_1, &
\varrho^{(1)}_2 = \tau^{(1)}_2, &
\varrho^{(1)}_3 = \tau^{(1)}_3, &
\varrho^{(1)}_4 = \tau^{(1)}_4, \\ \\
\varrho^{(2)}_1 = \tau^{(2)}_1, & 
\varrho^{(2)}_2 = \tau^{(2)}_2, &
\varrho^{(2)}_3 = \tau^{(1)}_1\oplus\tau^{(1)}_2, &
\varrho^{(2)}_4 = \tau^{(1)}_3\oplus\tau^{(1)}_4, \\ \\
\varrho^{(4)}_1 = \tau^{(4)}_1, & 
\varrho^{(4)}_2 = \tau^{(4)}_2, &
\varrho^{(4)}_3 = \tau^{(4)}_3, &
\varrho^{(4)}_4 = \tau^{(4)}_4, \\ \\
\varrho^{(4)}_5 = \tau^{(2)}_1\oplus\tau^{(2)}_2, & & & \\ \\
\varrho^{(6)}_1 = \tau^{(6)}_1, &
\varrho^{(6)}_2 = \tau^{(6)}_2, & &  \\ \\
\varrho^{(8)}_1 = \tau^{(8)}_1, &
\varrho^{(8)}_2 = \tau^{(8)}_2, &
\varrho^{(8)}_3 = \tau^{(4)}_1\oplus\tau^{(4)}_2, &
\varrho^{(8)}_4 = \tau^{(4)}_3\oplus\tau^{(4)}_4, \\ \\
\varrho^{(9)}_1 = \tau^{(3)}_1\oplus\tau^{(6)}_3, & 
\varrho^{(9)}_2 = \tau^{(3)}_2\oplus\tau^{(6)}_3, &
\varrho^{(9)}_3 = \tau^{(3)}_3\oplus\tau^{(6)}_4, &
\varrho^{(9)}_4 = \tau^{(3)}_4\oplus\tau^{(6)}_4, \\ \\
\varrho^{(12)}_1 = \tau^{(6)}_1\oplus\tau^{(6)}_2, & & & \\ \\
\varrho^{(16)}_1 = \tau^{(8)}_1\oplus\tau^{(8)}_2. & & & 
\end{array}
\eea

Table \ref{o4decomp} together with the above list shows how the irreducible $\Wc$ components fuse on the
16-cell honeycomb into irreducible $\W$ representations and will be the basis of all of the physical
consequences in the following sections.

\section{Rest frame kinematics}
\label{restframe}

In the subsequent three sections we will discuss the renormalization structure of 
polarized and unpolarized matrix elements. Generally, for an unpolarized matrix element of an operator in
some representation $\rho$ of the symmetry group let us write
\bea
{\cal M}_a(p) = \langle p | {\cal M}_a | p \rangle,
\label{mop}
\eea
where $a = 1,\ldots,\dim\rho$. Apart from the renormalization structure, it is also important to know
which operator and representation $\rho$ can be used in the rest frame, i.e. which one can have a 
non-zero matrix element for $p=(0,0,0,E)$. It is instructive to first discuss the continuum with symmetry
group $G = O(4)$, then the cubic and 16-cell lattices with $G=W(B_4)$ and $G=W(F_4)$, respectively. Let
$H \subset G$ denote the little group of $p = (0,0,0,E)$. Then matrix elements of the type (\ref{mop})
are fully determined by their value on the orbit of $p$, which is $G/H$, since we have,
\bea
{\cal M}_a( g\cdot p ) = \rho_{ab} {\cal M}_b(p).
\eea
Functions on $G/H$ carry a representation of $G$, let us call it ${\cal F}_G$, which decomposes into irreducibles.
In the continuum, of course, we have $H = O(3)$ and $G/H = O(4)/O(3) = S^3$ and the functions on $S^3$
decompose into spherical harmonics which carry the irreducibles $H_n$,
\bea
{\cal F}_{O(4)} = H_0 \oplus H_1 \oplus H_2 \oplus H_3 \oplus \ldots
\label{ghcont}
\eea
On the lattice the orbits are of course finite sets and in both cases the little group is $O_h$ of 48 elements. 
The difference between the two
lattices appears in the size of the orbit, which consists of the nearest neighbors (short roots). There are 8 on the
cubic lattice and 24 on the 16-cell honeycomb, as we have seen in section \ref{lattices}. Hence we have
$G/H = W(B_4) / O_h$ and $G/H = W(F_4) / O_h$ in the two cases, matching $8 = 384/48$ and $24 = 1152/48$,
respectively. Hence functions on the orbits carry an 8- and a 24-dimensional representation of the symmetry
groups in the two cases. These can be decomposed into irreducibles of the lattice symmetry groups, 
analogously to (\ref{ghcont}),
\bea
\label{ghlat}
{\cal F}_{W(B_4)} &=& \tau^{(1)}_1 \oplus \tau^{(3)}_1 \oplus \tau^{(4)}_1  \\
{\cal F}_{W(F_4)} &=& \varrho^{(1)}_1 \oplus \varrho^{(2)}_3 \oplus \varrho^{(4)}_1 \oplus
\varrho^{(8)}_3 \oplus \varrho^{(9)}_1 \nn\,,
\eea
matching $8 = 1 + 3 + 4$ and $24 = 1 + 2 + 4 + 8 + 9$, respectively. The decomposition
directly shows which irreducible lattice symmetry group representations can have a non-zero
unpolarized matrix element. Using table \ref{o4decomp} one may furthermore trace which component
in (\ref{ghlat}) originates from which $H_n$ continuum component of (\ref{ghcont}). As we will see in the
next three sections, the fact that two more irreducible components appear above on the 16-cell honeycomb
than on the cubic lattice, will lead to more matrix elements being measurable at rest, which is a
significant improvement in practice.

A similar analysis applies for polarized matrix elements. Let us introduce the spin 4-vector $S$ such that
$S\cdot p = 0$ and $S^2 = - E^2$. At rest, $p=(0,0,0,E)$, let us choose the polarization along the
$z$-direction, $S = (0,0,E,0)$. Generally for a polarized matrix element we have,
\bea
{\cal M}_a(p,S) = S_\mu {\cal M}_{\mu a}(p) = \langle p, S | {\cal M}_a | p, S \rangle,
\label{mops}
\eea
because it is linear in the spin. Hence the relevant representation space is,
\bea
{\cal S}_G = \det \otimes R \otimes {\cal F}_G \ominus {\cal F}_G,
\eea
where the subtraction of ${\cal F}_G$ corresponds to the removal of the factor $S\cdot p = 0$.
In the continuum, from (\ref{ghcont}) we get,
\bea
R \otimes H_0 &=& H_1 \nn \\
R \otimes H_n &=& H_{n+1} \oplus H_{n-1} \oplus M_{n+1}, \qquad \qquad n \geq 1.\nn
\eea
The removal of ${\cal F}_G$ corresponds to the $H_{n-1}$ factor
above, because that carries $n-1$ indices out of the total $n$. Tensoring with $\det$ then gives,
after taking into account that $\det \otimes M_{n+1} = M_{n+1}$ and the removal of the $H_{n-1}$ factor,
\bea
{\cal S}_{O(4)} = \mathop{\oplus}_{n=1}^\infty \left(  \det \otimes H_n \oplus M_{n+1} \right).
\eea
Both terms have transparent meanings, $\det\otimes H_n$ corresponds to
$\gamma_5$-inserted twist-2 and $M_{n+1}$ to twist-3 operators; see next sections.

On the lattice the starting point is (\ref{ghlat}) which needs to be tensored with $\det \otimes R$. On
the cubic lattice we have $R = \tau^{(4)}_1$ and,
\bea
\tau^{(4)}_1 \otimes \tau^{(1)}_1 &=& \tau^{(4)}_1 \nn \\
\tau^{(4)}_1 \otimes \tau^{(3)}_1 &=& \tau^{(4)}_1 \oplus \tau^{(8)}_1 \\
\tau^{(4)}_1 \otimes \tau^{(4)}_1 &=& \tau^{(1)}_1 \oplus \tau^{(3)}_1 \oplus \tau^{(6)}_1 \oplus \tau^{(6)}_3 \nn 
\eea
where in the second line $\tau^{(4)}_1$ and in the third line $\tau^{(1)}_1 \oplus \tau^{(3)}_1$ need
to be removed corresponding to the 3 components in (\ref{ghlat}). 
Tensoring further with $\det$ we then obtain the relevant
decomposition for non-zero polarized matrix elements,
\bea
\label{polcubic}
{\cal S}_{W(B_4)} = \tau^{(4)}_4 \oplus \tau^{(6)}_1 \oplus \tau^{(6)}_4 \oplus \tau^{(8)}_2\;.
\eea
On the 16-cell honeycomb we proceed similarly with $R = \varrho^{(4)}_1$ and from (\ref{ghlat}),
\bea
\varrho^{(4)}_1\otimes\varrho^{(1)}_1 &=& \varrho^{(4)}_1 \nn \\ 
\varrho^{(4)}_1\otimes\varrho^{(4)}_1 &=& \varrho^{(1)}_1\oplus\varrho^{(6)}_1\oplus\varrho^{(9)}_1 \nn \\
\varrho^{(4)}_1\otimes\varrho^{(2)}_3 &=& \varrho^{(8)}_3 \\
\varrho^{(4)}_1\otimes\varrho^{(9)}_1 &=& \varrho^{(4)}_1\oplus\varrho^{(8)}_1
\oplus\varrho^{(8)}_3 \oplus\varrho^{(16)}_1 \nn \\
\varrho^{(4)}_1\otimes\varrho^{(8)}_3 &=& \varrho^{(2)}_3\oplus\varrho^{(9)}_1\oplus\varrho^{(9)}_2
\oplus\varrho^{(12)}_1,  \nn
\eea
from which all 5 components of (\ref{ghlat}) need to be removed once, then tensoring with $\det$ gives,
\bea
\label{pol16}
{\cal S}_{W(F_4)} = \varrho^{(4)}_4\oplus\varrho^{(6)}_1\oplus\varrho^{(8)}_2 \oplus \varrho^{(8)}_4 \oplus\varrho^{(9)}_3 
\oplus \varrho^{(9)}_4 \oplus\varrho^{(12)}_1 \oplus \varrho^{(16)}_1.
\eea
Only these irreducible components can have a non-zero polarized matrix element at rest, exactly twice as many as
on the cubic lattice. 

\section{Unpolarized twist-2 matrix elements and moments of PDFs}
\label{twist2sec}

In this section the consequences of the representation theory in section \ref{irrep16} 
on unpolarized twist-2 operators will be explored. For the polarized case, see the next section. 
The totally symmetric, traceless and flavor non-singlet operators
\bea
\label{twist2}
{\cal O}_{\mu_1 \ldots \mu_n} = \left. {\bar \psi} \gamma_{\mu_1} i D_{\mu_2} \ldots i D_{\mu_n} \psi
\right|_{H_n},
\eea
form $H_n$ and are of mass dimension $n+2$ and spin $n$, so-called twist-2 operators, and they determine the
moments of parton distribution functions. The notation $\left. \right|_{H_n}$, or another representation
in the coming sections,
means that a projection onto that representation is implied in the indices, in this case it is made
totally symmetric and traceless. With a nucleon of momentum $p$, averaged over spin, we have,
\bea
\langle p |  {\cal O}_{\mu_1 \ldots \mu_n} | p \rangle = 2 \langle x^{n-1} \rangle \left. p_{\mu_1} \ldots
p_{\mu_n} \right|_{H_n}
\label{kin}
\eea
and lattice simulations determine the left hand side, inferring the moments $v_n = \langle x^{n-1} \rangle$ in
Bjorken-$x$. In (\ref{twist2}), and throughout this paper, we use the convention $D_\mu =
\tfrac12(\overrightarrow{D}_\mu-\overleftarrow{D}_\mu)$ for the symmetric left-right covariant derivative, 
so the operators ${\cal O}$ have definite charge conjugation eigenvalues, $C = (-1)^n$.

While the totally symmetric traceless tensors form an irreducible representation of $O(4)$, they decompose
under the discrete lattice symmetry groups. Mixing with lower dimensional operators and power
divergences may occur complicating renormalization and as we will see the details depend crucially on the
symmetry group $\Wc$ vs. $\W$.

Operators of the same dimension mix at most logarithmically and operators of lower dimensions mix with
power divergent coefficients. The lower dimensional operators may come in the form,
\bea
{\cal O}^\Gamma_{\mu_1 \ldots \mu_k} = {\bar \psi} \Gamma D_{\mu_1} \ldots D_{\mu_k} \psi 
\label{low}
\eea
with $\Gamma$ any one of $1,\; \gamma_5,\; \gamma_\mu,\; \gamma_5 \gamma_\mu,\; \sigma_{\mu\nu}$.
The mass dimension of (\ref{low}) is $k+3$. 
Total derivative operators will enter only in off-forward matrix elements and will be discussed in section \ref{gff}.
Let us denote the space of operators of the type (\ref{low})
by $\Omega(k)$, then we have,
\bea
\Omega(k) = ( 1 \,\oplus\, \det \,\oplus\, R \,\oplus\, \det \otimes R \,\oplus\, A ) 
\otimes R \otimes \cdots \otimes R,
\eea
with $k$ tensor products of $R$ at the end. Power divergent coefficients may appear
in the renormalization of twist-2 operators (\ref{twist2}) in a given irreducible representation of
the lattice symmetry group if lower dimensional ones exist in the same
representation. Hence the potentially dangerous arrangement is $k < n - 1$. If more than one lattice operator
exists at $k=n-1$, so with the same mass dimension, that will result in a logarithmic mixing matrix, as
opposed to the operator renormalizing logarithmically with just a single $Z$-factor. These are of course
less severe than the power divergences, but may still complicate the calculation.

Three other considerations are to be
discussed for a full analysis. First, other symmetries like charge conjugation, $C$, or chirality may forbid
mixing with lower dimensional operators which would otherwise be allowed by the lattice symmetry group.
Since our focus will be Wilson fermions, the constraint from chirality will not be imposed, but charge
conjugation will. The twist-2 operators in $H_n$ have $C=(-1)^n$.
Let us introduce $\Omega^\pm(k)$ as the $C$-even and $C$-odd part of
$\Omega(k)$. It is easy to see that the dimensions are,
\bea
\label{dimomega}
\dim\Omega(k) &=& 2^{2k+4}  \\
\dim\Omega^\pm(k) &=& 2^{2k+3} \pm (-1)^{k+1} 2^{2\left[\frac{k+1}{2}\right] + 1}. \nn
\eea
Second, equations of motion operators can be discarded in on-shell matrix elements, 
hence let us introduce
$\Omega_{os}(k)$ and $\Omega^\pm_{os}(k)$ for the on-shell spaces where equations of motion operators are
subtracted. Note that there are no equations of motion operators at $k=0$,
so $\Omega_{os}(0) = \Omega(0)$ and similarly $\Omega^\pm_{os}(0) = \Omega^\pm(0)$. 
The dimensions for $k=1$ are,
\bea
\dim\Omega_{os}(1) = 34 \qquad\qquad \dim\Omega^+_{os}(1) = 25  \qquad\qquad \dim\Omega^-_{os}(1) = 9\;
\label{dimomega1}
\eea
and for $k>1$,
\bea
\label{dimomegak}
\dim\Omega_{os}(k) &=&  9 \cdot 2^{2k} \\
\dim\Omega^\pm_{os}(k) &=& 9 \cdot 2^{2k-1} \pm 3 (-1)^{k+1} 2^{2\left[\frac{k+1}{2}\right]-1}. \nn
\eea
The expressions (\ref{dimomega}), (\ref{dimomega1}) and (\ref{dimomegak}) are useful to check that for all
decompositions into irreducibles we encounter, the dimensions match on the left and right hand side.
Third, even if no power divergent mixing occurs, the required momenta may need one or more non-zero
spatial components, which introduces signal-to-noise
problems in practice. This may happen if all polynomials in the representation in question vanish in the
rest frame $p=(0,0,0,E)$, which aspect was analyzed in detail in section \ref{restframe}.

Before going to the two lattice discretizations, it is instructive to see the continuum structure. Each
$\Omega^\pm(k), \Omega^\pm_{os}(k)$ 
decomposes into $O(4)$ irreducible representations, the multiplicity of each is listed in table
\ref{o4cont}. Each $H_n$ appears exactly once in $\Omega^{(-1)^n}_{os}(n-1)$,  
hence the renormalization is always logarithmic in the continuum.

\begin{table}
\centering
\renewcommand{\arraystretch}{1.3}
\setlength{\tabcolsep}{4pt}
\begin{tabular}{|c|c||c||c|c||c|c||c|c|}
\hline
     &  & $\Omega^\pm(0)$ & $\Omega^\pm(1)$ & $\Omega^\pm_{os}(1)$ & $\Omega^\pm(2)$ & $\Omega^\pm_{os}(2)$ &
     $\Omega^\pm(3)$ & $\Omega^\pm_{os}(3)$ \\
\hline
\hline
$(0, 0)^+$ & 1                                           & 1, 0 & 1, 0 & 0, 0 & 2, 0 & 1, 0 & 3, 1 & 1, 0 \\
\hline                                                                                                  
$(0, 0)^-$ & $\det$                                      & 1, 0 & 0, 1 & 0, 0 & 2, 0 & 1, 0 & 1, 3 & 0, 1 \\
\hline                                                                                                      
$(\tfrac12, \tfrac12)^+$ & $R$                           & 0, 1 & 1, 1 & 0, 0 & 1, 3 & 0, 1 & 5, 5 & 2, 2 \\
\hline                                                                                                      
$(\tfrac12, \tfrac12)^-$ & $\det\otimes R$               & 1, 0 & 1, 1 & 0, 0 & 3, 1 & 1, 0 & 5, 5 & 2, 2 \\
\hline                                                                                                      
$(1, 0)\oplus(0, 1)$ & $A$                               & 0, 1 & 1, 1 & 0, 0 & 1, 4 & 0, 2 & 6, 6 & 2, 2 \\
\hline                                                                                                      
$(1, 1)^+$ & $H_2$                                       & 0, 0 & 1, 0 & 1, 0 & 2, 1 & 1, 0 & 6, 3 & 3, 1 \\
\hline                                                                                                      
$(1, 1)^-$ & $\det\otimes H_2$                           & 0, 0 & 0, 1 & 0, 1 & 2, 1 & 1, 0 & 3, 6 & 1, 3 \\
\hline                                                                                                      
$(\tfrac32, \tfrac12)\oplus(\tfrac12, \tfrac32)$ & $M_3$ & 0, 0 & 1, 0 & 1, 0 & 2, 2 & 1, 1 & 7, 6 & 3, 3 \\
\hline                                                                                                      
$(\tfrac32, \tfrac32)^+$ & $H_3$                         & 0, 0 & 0, 0 & 0, 0 & 0, 1 & 0, 1 & 2, 2 & 1, 1 \\
\hline                                                                                                      
$(\tfrac32, \tfrac32)^-$ & $\det\otimes H_3$             & 0, 0 & 0, 0 & 0, 0 & 1, 0 & 1, 0 & 2, 2 & 1, 1 \\
\hline                                                                                                      
$(2, 0)\oplus(0, 2)$ & $W$                               & 0, 0 & 0, 0 & 0, 0 & 1, 0 & 1, 0 & 2, 2 & 1, 1 \\
\hline                                                                                                      
$(2, 1)\oplus(1, 2)$ & $M_4$                             & 0, 0 & 0, 0 & 0, 0 & 0, 1 & 0, 1 & 3, 3 & 2, 2 \\
\hline                                                                                                      
$(2, 2)^+$ & $H_4$                                       & 0, 0 & 0, 0 & 0, 0 & 0, 0 & 0, 0 & 1, 0 & 1, 0 \\
\hline                                                                                                      
$(2, 2)^-$ & $\det\otimes H_4$                           & 0, 0 & 0, 0 & 0, 0 & 0, 0 & 0, 0 & 0, 1 & 0, 1 \\
\hline                                                                                                      
$(\tfrac52, \tfrac12)\oplus(\tfrac12, \tfrac52)$ &       & 0, 0 & 0, 0 & 0, 0 & 0, 0 & 0, 0 & 1, 1 & 1, 1 \\
\hline                                                                                                      
$(\tfrac52, \tfrac32)\oplus(\tfrac32, \tfrac52)$ & $M_5$ & 0, 0 & 0, 0 & 0, 0 & 0, 0 & 0, 0 & 1, 0 & 1, 0 \\
\hline
\end{tabular}
    \caption{Multiplicity of $O(4)$ representations, and their labels, in $\Omega^\pm(k)$
    and the on-shell $\Omega^\pm_{os}(k)$. Each multiplicity is split into $C$-even and $C$-odd
    multiplicities. We have $\Omega(0) = \Omega_{os}(0)$.}
\label{o4cont}
\end{table}

\subsection{Cubic lattice}

All of the statements below are well-established, we simply list them to put the 16-cell honeycomb
results, which are new, into perspective.

How each of the $\Omega^\pm(k)$ and $\Omega^\pm_{os}(k)$ spaces decomposes under $\Wc$ is listed in table
\ref{omegatau},
which lists the multiplicities of each of the 20 irreducible representations of $\Wc$.
A useful cross-check is the matching of dimensions, which is seen to hold for all columns.
The decomposition of the $H_n$ spaces under $\Wc$ is listed, among all the others, in table \ref{o4decomp}.
Let us infer the mixing patterns from these two tables.

\begin{table}
\renewcommand{\arraystretch}{1.3}
\setlength{\tabcolsep}{4pt}
\centering
\begin{tabular}[t]{|c||c||c|c||c|c||c|c|}
\hline
    $\Wc$ & $\Omega^\pm(0)$ & $\Omega^\pm(1)$ & $\Omega^\pm_{os}(1)$ & $\Omega^\pm(2)$ & $\Omega^\pm_{os}(2)$ & $\Omega^\pm(3)$ & $\Omega^\pm_{os}(3)$ \\
\hline
\hline
$\tau^{(1)}_1$ & 1, 0 & 1, 0 & 0, 0 & 2, 0 & 1, 0 & 4, 1 & 2, 0 \\
\hline
$\tau^{(1)}_2$ & 0, 0 & 0, 0 & 0, 0 & 0, 0 & 0, 0 & 1, 0 & 1, 0 \\
\hline
$\tau^{(1)}_3$ & 0, 0 & 0, 0 & 0, 0 & 0, 0 & 0, 0 & 0, 1 & 0, 1 \\
\hline
$\tau^{(1)}_4$ & 1, 0 & 0, 1 & 0, 0 & 2, 0 & 1, 0 & 1, 4 & 0, 2 \\
\hline
$\tau^{(2)}_1$ & 0, 0 & 0, 0 & 0, 0 & 1, 0 & 1, 0 & 3, 2 & 2, 1 \\
\hline
$\tau^{(2)}_2$ & 0, 0 & 0, 0 & 0, 0 & 1, 0 & 1, 0 & 2, 3 & 1, 2 \\
\hline
$\tau^{(3)}_1$ & 0, 0 & 1, 0 & 1, 0 & 2, 1 & 1, 0 & 7, 3 & 4, 1 \\
\hline
$\tau^{(3)}_2$ & 0, 0 & 0, 0 & 0, 0 & 0, 1 & 0, 1 & 3, 3 & 2, 2 \\
\hline
$\tau^{(3)}_3$ & 0, 0 & 0, 0 & 0, 0 & 0, 1 & 0, 1 & 3, 3 & 2, 2 \\
\hline
$\tau^{(3)}_4$ & 0, 0 & 0, 1 & 0, 1 & 2, 1 & 1, 0 & 3, 7 & 1, 4 \\
\hline
$\tau^{(4)}_1$ & 0, 1 & 1, 1 & 0, 0 & 1, 4 & 0, 2 & 8, 7 & 4, 3 \\
\hline
$\tau^{(4)}_2$ & 0, 0 & 0, 0 & 0, 0 & 0, 1 & 0, 1 & 4, 3 & 3, 2 \\
\hline
$\tau^{(4)}_3$ & 0, 0 & 0, 0 & 0, 0 & 1, 0 & 1, 0 & 4, 3 & 3, 2 \\
\hline
$\tau^{(4)}_4$ & 1, 0 & 1, 1 & 0, 0 & 4, 1 & 2, 0 & 8, 7 & 4, 3 \\
\hline
$\tau^{(6)}_1$ & 0, 1 & 1, 1 & 0, 0 & 1, 5 & 0, 3 & 10, 10 & 5, 5 \\
\hline
$\tau^{(6)}_2$ & 0, 0 & 0, 0 & 0, 0 & 1, 1 & 1, 1 & 6, 6 & 4, 4 \\
\hline
$\tau^{(6)}_3$ & 0, 0 & 1, 0 & 1, 0 & 2, 2 & 1, 1 & 10, 6 & 6, 3 \\
\hline
$\tau^{(6)}_4$ & 0, 0 & 0, 1 & 0, 1 & 2, 2 & 1, 1 & 6, 10 & 3, 6 \\
\hline
$\tau^{(8)}_1$ & 0, 0 & 1, 0 & 1, 0 & 2, 3 & 1, 2 & 12, 9 & 7, 5 \\
\hline
$\tau^{(8)}_2$ & 0, 0 & 1, 0 & 1, 0 & 3, 2 & 2, 1 & 12, 9 & 7, 5 \\
\hline
\end{tabular}
    \caption{Multiplicity of $\Wc$ representations in $\Omega^\pm(k)$
    and the on-shell $\Omega^\pm_{os}(k)$. Each multiplicity is split into $C$-even and $C$-odd
    multiplicities. We have $\Omega(0) = \Omega_{os}(0)$.}
    \label{omegatau}
\end{table}

\begin{table}
\renewcommand{\arraystretch}{1.3}
\setlength{\tabcolsep}{4pt}
\centering
\begin{tabular}[t]{|c||c||c|c||c|c||c|c|}
\hline
    $\W$ & $\Omega^\pm(0)$ & $\Omega^\pm(1)$ & $\Omega^\pm_{os}(1)$ & $\Omega^\pm(2)$ & $\Omega^\pm_{os}(2)$ & $\Omega^\pm(3)$ & $\Omega^\pm_{os}(3)$ \\
\hline
\hline
$\varrho^{(1)}_{1}$ & 1, 0 & 1, 0 & 0, 0 & 2, 0 & 1, 0 & 3, 1 & 1, 0 \\
\hline
$\varrho^{(1)}_{2}$ & 0, 0 & 0, 0 & 0, 0 & 0, 0 & 0, 0 & 0, 0 & 0, 0 \\
\hline
$\varrho^{(1)}_{3}$ & 0, 0 & 0, 0 & 0, 0 & 0, 0 & 0, 0 & 0, 0 & 0, 0 \\
\hline
$\varrho^{(1)}_{4}$ & 1, 0 & 0, 1 & 0, 0 & 2, 0 & 1, 0 & 1, 3 & 0, 1 \\
\hline
$\varrho^{(2)}_{1}$ & 0, 0 & 0, 0 & 0, 0 & 0, 0 & 0, 0 & 1, 0 & 1, 0 \\
\hline
$\varrho^{(2)}_{2}$ & 0, 0 & 0, 0 & 0, 0 & 0, 0 & 0, 0 & 0, 1 & 0, 1 \\
\hline
$\varrho^{(2)}_{3}$ & 0, 0 & 0, 0 & 0, 0 & 0, 0 & 0, 0 & 1, 0 & 1, 0 \\
\hline
$\varrho^{(2)}_{4}$ & 0, 0 & 0, 0 & 0, 0 & 0, 0 & 0, 0 & 0, 1 & 0, 1 \\
\hline
$\varrho^{(4)}_{1}$ & 0, 1 & 1, 1 & 0, 0 & 1, 3 & 0, 1 & 5, 5 & 2, 2 \\
\hline
$\varrho^{(4)}_{2}$ & 0, 0 & 0, 0 & 0, 0 & 0, 0 & 0, 0 & 1, 1 & 1, 1 \\
\hline
$\varrho^{(4)}_{3}$ & 0, 0 & 0, 0 & 0, 0 & 0, 0 & 0, 0 & 1, 1 & 1, 1 \\
\hline
$\varrho^{(4)}_{4}$ & 1, 0 & 1, 1 & 0, 0 & 3, 1 & 1, 0 & 5, 5 & 2, 2 \\
\hline
$\varrho^{(4)}_{5}$ & 0, 0 & 0, 0 & 0, 0 & 1, 0 & 1, 0 & 2, 2 & 1, 1 \\
\hline
$\varrho^{(6)}_{1}$ & 0, 1 & 1, 1 & 0, 0 & 1, 4 & 0, 2 & 6, 6 & 2, 2 \\
\hline
$\varrho^{(6)}_{2}$ & 0, 0 & 0, 0 & 0, 0 & 1, 0 & 1, 0 & 2, 2 & 1, 1 \\
\hline
$\varrho^{(8)}_{1}$ & 0, 0 & 0, 0 & 0, 0 & 0, 1 & 0, 1 & 3, 2 & 2, 1 \\
\hline
$\varrho^{(8)}_{2}$ & 0, 0 & 0, 0 & 0, 0 & 1, 0 & 1, 0 & 3, 2 & 2, 1 \\
\hline
$\varrho^{(8)}_{3}$ & 0, 0 & 0, 0 & 0, 0 & 0, 1 & 0, 1 & 3, 2 & 2, 1 \\
\hline
$\varrho^{(8)}_{4}$ & 0, 0 & 0, 0 & 0, 0 & 1, 0 & 1, 0 & 3, 2 & 2, 1 \\
\hline
$\varrho^{(9)}_{1}$ & 0, 0 & 1, 0 & 1, 0 & 2, 1 & 1, 0 & 7, 3 & 4, 1 \\
\hline
$\varrho^{(9)}_{2}$ & 0, 0 & 0, 0 & 0, 0 & 0, 1 & 0, 1 & 3, 3 & 2, 2 \\
\hline
$\varrho^{(9)}_{3}$ & 0, 0 & 0, 0 & 0, 0 & 0, 1 & 0, 1 & 3, 3 & 2, 2 \\
\hline
$\varrho^{(9)}_{4}$ & 0, 0 & 0, 1 & 0, 1 & 2, 1 & 1, 0 & 3, 7 & 1, 4 \\
\hline
$\varrho^{(12)}_{1}$ & 0, 0 & 0, 0 & 0, 0 & 0, 1 & 0, 1 & 4, 4 & 3, 3 \\
\hline
$\varrho^{(16)}_{1}$ & 0, 0 & 1, 0 & 1, 0 & 2, 2 & 1, 1 & 9, 7 & 5, 4 \\
\hline
\end{tabular}
    \caption{Multiplicity of $\W$ representations in $\Omega^\pm(k)$
    and the on-shell $\Omega^\pm_{os}(k)$. Each multiplicity is split into $C$-even and $C$-odd
    multiplicities. We have $\Omega(0) = \Omega_{os}(0)$.}
    \label{omegarho}
\end{table}

In $H_2$, relevant for $v_2 = \langle x \rangle$ which has $C=1$, 
there are two components, $\tau^{(3)}_1$ and $\tau^{(6)}_3$ and 
neither appears in $\Omega^+_{os}(0)$ and each appears only once in $\Omega^+_{os}(1)$, 
hence we arrive at the familiar conclusion that $v_2$ renormalizes only multiplicatively.
The kinematics are different in the two cases, all polynomials vanish at rest in $\tau^{(6)}_3$, at least
one spatial momentum component needs to be non-zero, while $\tau^{(3)}_1$ is usable at rest, since it
appears in (\ref{ghlat}), making it
the preferred channel on the cubic lattice. Lattice determinations of $v_2$ on the cubic lattice include
\cite{Gockeler:2004wp,Orginos:2005uy,Lin:2008uz,Aoki:2010xg,LHPC:2010jcs,Bali:2014nma,Mondal:2020ela,Rodekamp:2023wpe},
see also \cite{Bar:2016jof} for the contamination from nucleon-pion states.

In $H_3$, relevant for $v_3 = \langle x^2 \rangle$, we see from table \ref{o4decomp} 
that there are three components, $\tau^{(4)}_1, \tau^{(4)}_2$ and $\tau^{(8)}_1$. We have
$C=-1$, hence appearance in $\Omega^-_{os}(k)$ is the relevant question. The fate of all three components is,
\begin{itemize}
    \item $\tau^{(4)}_1$, appears in $\Omega^-_{os}(0)$ leading to $1/a^2$ divergences and twice in
        $\Omega^-_{os}(2)$ leading to a $2 \times 2$ logarithmic mixing matrix.
    \item $\tau^{(4)}_2$, only appears once in $\Omega^-_{os}(2)$ hence there are no power divergences and
        no enlarged logarithmic mixing.
    \item $\tau^{(8)}_1$, only appears in $\Omega^-_{os}(2)$, but twice, so there are no power divergences
        but the logarithmic mixing matrix is $2\times 2$.
\end{itemize}
Out of the two candidates without power divergences, 
$\tau^{(4)}_2$ and $\tau^{(8)}_1$, both are kinematically unfavorable, in the
rest frame all polynomials vanish; they do not appear in (\ref{ghlat}). 
At least two spatial momentum components need to be non-zero in
$\tau^{(4)}_2$ and at least one in $\tau^{(8)}_1$. Hence, the best case scenario is $\tau^{(8)}_1$,
although the one non-zero spatial momentum component introduces a bad signal-to-noise problem and also
the logarithmic mixing matrix is enlarged to $2\times 2$.

In $H_4$, which would be needed for $v_4 = \langle x^3 \rangle$ and has $C=1$, there are seven components 
and their renormalization structure is listed for all of them below,
\begin{itemize}
    \item $\tau^{(1)}_1$, this is the trivial representation and has a component in both
        $\Omega^+_{os}(0)$ and $\Omega^+_{os}(2)$ leading to $1/a^3$ and $1/a$ divergences. It also
        appears twice in $\Omega^+_{os}(3)$, resulting in a $2 \times 2$ logarithmic mixing matrix.
    \item $\tau^{(1)}_2$, behaves exactly the same as in the continuum. 
    \item $\tau^{(2)}_1$, appears in $\Omega^+_{os}(2)$, generating a $1/a$ power divergence. It also
        appears twice in $\Omega^+_{os}(3)$ resulting in a $2\times 2$ logarithmic mixing matrix.
    \item $\tau^{(3)}_1$, appears in both $\Omega^+_{os}(1)$ and $\Omega^+_{os}(2)$ 
        generating $1/a^2,\; 1/a$ divergences. The logarithmic mixing matrix becomes $4\times 4$ since it
        appears four times in $\Omega^+_{os}(3)$.
    \item $\tau^{(6)}_1$, only appears in $\Omega^+_{os}(3)$, five times, generating no power divergences
        but a $5 \times 5$ logarithmic mixing matrix.
    \item $\tau^{(6)}_2$, appears in $\Omega^+_{os}(2)$, generating a $1/a$ power divergence and it also
        appears in $\Omega^+_{os}(3)$ four times leading to a $4 \times 4$ logarithmic mixing matrix.
    \item $\tau^{(6)}_3$, same power divergences as in $\tau^{(3)}_1$, but the logarithmic mixing matrix
        is enlarged to $6 \times 6$.
\end{itemize}
The most promising candidates from the above list are then $\tau^{(1)}_2$ and $\tau^{(6)}_1$, both free
from power divergences. The kinematics are unfavorable again, the polynomial in $\tau^{(1)}_2$ is
$p_1p_2p_3p_4$ so all spatial momentum components need to be non-zero, making this channel prohibitive.
The polynomials in $\tau^{(6)}_1$ are $p_\mu p_\nu (p_\mu^2 - p_\nu^2)$ for $\mu < \nu$ which vanish in
the rest frame, at least 1 spatial momentum component needs to be non-zero, but there is of course the
problem of a $5 \times 5$ mixing matrix of same dimension operators too. Nevertheless, the nucleon $v_3$ and
$v_4$ were computed on the cubic lattice at non-zero momentum in \cite{Gockeler:2004wp}, and for the pion
and kaon moments up to $v_4$ were obtained in \cite{Alexandrou:2020gxs,Alexandrou:2021mmi,Alexandrou:2026nsl}.

\subsection{16-cell honeycomb}

Now the same analysis for the 16-cell honeycomb proceeds completely analogously.
How each of the $\Omega^\pm_{os}(k)$ spaces decomposes under $\W$ is listed in table \ref{omegarho}
which lists the multiplicities of each of the 25 irreducible representations of $\W$.
Again the useful cross-check of matching dimensions is clear along all columns.
The decomposition of $H_n$ into irreducibles of $\W$ is shown in table \ref{o4decomp}. 

In $H_2$, relevant for $v_2 = \langle x \rangle$, there is a single component, $\varrho^{(9)}_1$, which only
appears once in $\Omega^+_{os}(1)$, the same way it happens in the continuum, 
hence renormalization is logarithmic only, and the mixing matrix is not enlarged. The kinematics is
favorable, this representation appears in (\ref{ghlat}) hence is usable at rest.

In $H_3$, relevant for $v_3 = \langle x^2 \rangle$, there are two components, $\varrho^{(8)}_1$ and
$\varrho^{(8)}_3$, both behaving exactly the same as in the continuum again.
The first one, $\varrho^{(8)}_1$, is however
less useful, because in the rest frame all 8 polynomials vanish, at least one spatial component needs to
be non-zero. The component $\varrho^{(8)}_3$ appears in (\ref{ghlat}) hence can be measured in the rest frame, 
making it the most useful channel for $v_3 = \langle x^2 \rangle$.

There are four components in $H_4$, needed for $v_4 = \langle x^3 \rangle$, instead of seven as in the cubic
case. Their divergence structure, from table \ref{omegarho} is, 
\begin{itemize}
    \item $\varrho^{(2)}_1$, behaves exactly the same way as in the continuum.
    \item $\varrho^{(2)}_3$, also behaves exactly the same way as in the continuum.  
    \item $\varrho^{(9)}_1$, appears in both $\Omega^+_{os}(1)$ and $\Omega^+_{os}(2)$ 
        generating $1/a^2,\;1/a$ power divergences and also in $\Omega^+_{os}(3)$ four times, leading to
        a $4 \times 4$ logarithmic mixing matrix.
    \item $\varrho^{(12)}_1$, free from power divergences but appears three times in $\Omega^+_{os}(3)$
        so the logarithmic mixing matrix is enlarged to $3 \times 3$.
\end{itemize}
Except for $\varrho^{(9)}_1$, they are free from power divergences and the most promising candidates are
$\varrho^{(2)}_1$ and $\varrho^{(2)}_3$ because in these two cases the mixing matrix is not enlarged. In
the former, $\varrho^{(2)}_1$, the kinematics are unfavorable, all polynomials vanish at rest, at least
one spatial momentum component needs to be non-zero, while the latter, $\varrho^{(2)}_3$ appears in
(\ref{ghlat}) hence usable at rest,  making it the most useful channel.

For completeness we list all 8 polynomials for $\varrho^{(8)}_3$, which is the ideal channel for $\langle
x^2 \rangle$ below,
\bea
p_\mu \left( p_\mu^2 - \frac12 p^2 \right), \qquad \frac{p_1 p_2 p_3 p_4}{p_\mu},
\eea
where the second set of four vanishes in the rest frame, while one of the first four, with $\mu=4$, gives
$\tfrac12 E^3$ for $p=(0,0,0,E)$. Similarly, the two polynomials for $\varrho^{(2)}_3$, the ideal 
candidate for $v_4 = \langle x^3 \rangle$, are,
\bea
\sum_\mu p_\mu^4 - \frac12 (p^2)^2, \qquad p_1 p_2 p_3 p_4,
\eea
where the second one vanishes in the rest frame but the first one gives $\tfrac12 E^4$. 

Even on the 16-cell honeycomb no further moments, $v_n = \langle x^{n-1} \rangle$ with $n>4$, 
can be measured without power divergences, the best case contains $1/a^2$ for $n=5$.

\subsection{Flavor singlet and gluonic operators}
\label{fs2}

So far only flavor non-singlet bilinears (\ref{twist2}) were considered. In the continuum,
operators ${\cal O}^s$ of the form (\ref{twist2}) but with a flavor singlet structure,
with even $n$, mix with purely gluonic operators as well. 
Let us denote the gluonic $C=1$ twist-2 operators by,
\bea
{\cal O}^g_{\mu_1\ldots\mu_n} = \left. \tr F_{\mu_1 \nu} D_{\mu_2} \ldots D_{\mu_{n-1}} F_{\mu_n \nu} \right|_{H_n},
\eea
hence $({\cal O}^s, {\cal O}^g)$ mix for even $n$. For lattice determinations of the forward and off-forward
matrix elements of ${\cal O}^g$ on the cubic lattice see \cite{Detmold:2017oqb,Winter:2017bfs,Shanahan:2018pib}, and for the
gluon momentum fraction of the nucleon and the pion
\cite{QCDSF:2012mkm,Deka:2013zha,Alexandrou:2016ekb,Yang:2018bft,ExtendedTwistedMass:2021rdx,Wang:2021vqy,Fan:2022qve}. Since we are only
interested in $n < 5$, the only relevant cases are $n=2,4$. Let us denote the corresponding moments by
$\langle x^{n-1} \rangle_s$ and $\langle x^{n-1} \rangle_g$.

On the lattice potentially power divergent coefficients may arise, beyond the 
$\Omega^\pm_{os}(k)$ families we have seen already,
from the unit operator and from operators of the form $\tr( F D \ldots D F )$ with $k$ covariant derivatives. 
Since $n$ is even $k$ needs to be even too. The
mass dimension is $k+4$, hence at $n=2$ there cannot occur any power divergence from these. On both
lattices $\langle x \rangle_s$ and $\langle x \rangle_g$ can be measured with logarithmic $2 \times 2$
mixing that is present in the continuum as well. Thus in the cubic case there are two different mixing
matrices while on the 16-cell honeycomb there continues to be a single one. 

At $n=4$, relevant for $v_4$, only $k=0$
may generate power divergences. The $k=0$
space of operators is the symmetrized part of $A \otimes A$, $Sym^2( A )$, consisting of all possible 
index structures out of two interchangeable field strength tensors, $\tr( F_{\mu\nu} F_{\rho\sigma} )$. This 
reducible representation
is $21$ dimensional and decomposes under $O(4)$ and, using table \ref{o4decomp}, further under $\W$ as,
\bea
Sym^2(A) = 1 \oplus \det \oplus H_2 \oplus W = \varrho^{(1)}_1 \oplus \varrho^{(1)}_4 \oplus
\varrho^{(9)}_1 \oplus \varrho^{(4)}_5 \oplus \varrho^{(6)}_2,
\label{rep21}
\eea
matching $21 = 1 + 1 + 9 + 10 = 1 + 1 + 9 + 4 + 6$. 
We will not inspect $\Wc$ in detail because on the cubic lattice the flavor non-singlet third 
moment was already confirmed to be unfavorable, the situation stays the same in the 
flavor singlet or gluonic sectors. 
On the 16-cell honeycomb we have, from table \ref{o4decomp},
$H_4 = \varrho^{(2)}_1 \oplus \varrho^{(2)}_3 \oplus \varrho^{(9)}_1 \oplus \varrho^{(12)}_1$
and it does contain a common factor with (\ref{rep21}), which is $\varrho^{(9)}_1$, 
but this component was already eliminated from the list of
candidates because it generates a power divergence in the flavor non-singlet channel. a
There is a further $C=1$
potential gluonic operator of mass dimension 6, a 3-gluon operator, 
which even though does not mix in the continuum, might on the
16-cell honeycomb. This is the operator of the form $f_{abc} \tr( F^a F^b F^c )$, which is in the 
20 dimensional representation of totally anti-symmetric combination of $A \otimes A \otimes A$ which
under $O(4)$ decomposes as $1 \oplus \det \oplus H_2 \oplus \det\otimes H_2$. There is clearly no $H_4$
factor, hence no mixing in the continuum. 
Under $\W$ all four factors stay irreducible and we have $\varrho^{(1)}_1 \oplus \varrho^{(1)}_4
\oplus \varrho^{(9)}_1 \oplus \varrho^{(9)}_4$ and indeed $\varrho^{(2)}_3$ does not appear. 
In summary, $\varrho^{(2)}_3$, as we have seen does not appear in (\ref{rep21}), neither has a component
in the aforementioned 3-gluon operator, hence
renormalizes exactly as in the continuum and can be measured in the rest
frame not only in the flavor non-singlet sector but also in the gluonic or flavor singlet sector 
as well.

\subsection{Generalized form factors}
\label{gff}

The findings so far carry over to off-forward matrix elements too and the corresponding generalized form
factors. One aspect not present in forward matrix elements is that total derivative operators need to be
kept. Let us introduce $\partial\,\Omega(k)$ for the space of total derivative operators, which were not counted in
$\Omega(k)$ and $\partial\,\Omega^\pm(k)$ and $\partial\,\Omega^\pm_{os}(k)$, analogously. Operators in
$\partial\,\Omega(k)$ have $m$
partial derivatives on an element of $\Omega(k-m)$ with $m=1,\ldots,k$. The $m$ partial derivatives of
course are symmetric in the $m$ indices, hence
\bea
\partial\,\Omega(k) = \mathop{\oplus}_{m=1}^k Sym^m(R) \otimes \Omega(k-m),
\eea
summing all possible $m$ factors, where $Sym^m(R)$ is the $m$-index totally symmetric representation.
Consequently, the dimensions are the sum over all $m=1,\ldots,k$ subspaces and turn out to be,
\bea
\dim\partial\,\Omega(k) = \frac{8}{81} \left( 350 \cdot 4^k - 9 k^3 - 90 k^2 - 303 k - 350 \right).
\eea
In a completely analogous fashion to $\Omega^\pm(k)$ and $\Omega^\pm_{os}(k)$, the decomposition of
$\partial\,\Omega^\pm(k)$ and $\partial\,\Omega^\pm_{os}(k)$ into 
$O(4)$, $\Wc$ and $\W$ irreducibles can be found
in tables \ref{o4contd}, \ref{deltatau} and \ref{deltarho}, which will be used in what follows.

\begin{table}
\centering
\renewcommand{\arraystretch}{1.3}
\setlength{\tabcolsep}{4pt}
\begin{tabular}{|c|c||c|c||c|c||c|c|}
\hline
     &  & $\partial\,\Omega^\pm(1)$ & $\partial\,\Omega^\pm_{os}(1)$ & $\partial\,\Omega^\pm(2)$ & $\partial\,\Omega^\pm_{os}(2)$ & $\partial\,\Omega^\pm(3)$ & $\partial\,\Omega^\pm_{os}(3)$ \\
\hline
\hline
$(0, 0)^+$ & 1 & 0, 1 & 0, 1 & 2, 1 & 1, 0 & 3, 4 & 1, 2 \\
\hline
$(0, 0)^-$ & $\det$ & 1, 0 & 1, 0 & 2, 1 & 1, 0 & 4, 3 & 2, 1 \\
\hline
$(\tfrac12, \tfrac12)^+$ & $R$ & 1, 1 & 1, 1 & 3, 3 & 1, 2 & 9, 8 & 4, 3 \\
\hline
$(\tfrac12, \tfrac12)^-$ & $\det\otimes R$ & 1, 1 & 1, 1 & 3, 3 & 2, 1 & 9, 8 & 4, 3 \\
\hline
$(1, 0)\oplus(0, 1)$ & $A$ & 1, 1 & 1, 1 & 3, 4 & 1, 2 & 10, 10 & 4, 4 \\
\hline
$(1, 1)^+$ & $H_2$ & 0, 1 & 0, 1 & 3, 2 & 2, 1 & 7, 9 & 3, 5 \\
\hline
$(1, 1)^-$ & $\det\otimes H_2$ & 1, 0 & 1, 0 & 3, 2 & 2, 1 & 9, 7 & 5, 3 \\
\hline
$(\tfrac32, \tfrac12)\oplus(\tfrac12, \tfrac32)$ & $M_3$ & 0, 1 & 0, 1 & 3, 3 & 2, 2 & 11, 11 & 6, 5 \\
\hline
$(\tfrac32, \tfrac32)^+$ & $H_3$ & 0, 0 & 0, 0 & 1, 1 & 1, 1 & 5, 4 & 3, 2 \\
\hline
$(\tfrac32, \tfrac32)^-$ & $\det\otimes H_3$ & 0, 0 & 0, 0 & 1, 1 & 1, 1 & 5, 4 & 3, 2 \\
\hline
$(2, 0)\oplus(0, 2)$ & $W$ & 0, 0 & 0, 0 & 1, 0 & 1, 0 & 3, 3 & 2, 2 \\
\hline
$(2, 1)\oplus(1, 2)$ & $M_4$ & 0, 0 & 0, 0 & 1, 1 & 1, 1 & 6, 6 & 4, 4 \\
\hline
$(2, 2)^+$ & $H_4$ & 0, 0 & 0, 0 & 0, 0 & 0, 0 & 1, 2 & 1, 2 \\
\hline
$(2, 2)^-$ & $\det\otimes H_4$ & 0, 0 & 0, 0 & 0, 0 & 0, 0 & 2, 1 & 2, 1 \\
\hline
$(\tfrac52, \tfrac12)\oplus(\tfrac12, \tfrac52)$ &  & 0, 0 & 0, 0 & 0, 0 & 0, 0 & 2, 1 & 2, 1 \\
\hline
$(\tfrac52, \tfrac32)\oplus(\tfrac32, \tfrac52)$ & $M_5$ & 0, 0 & 0, 0 & 0, 0 & 0, 0 & 1, 2 & 1, 2 \\
\hline
\end{tabular}
    \caption{Multiplicity of $O(4)$ representations, and their labels, in the total derivative
    spaces $\partial\,\Omega^\pm(k)$ and the on-shell $\partial\,\Omega^\pm_{os}(k)$. Each multiplicity is split into
    $C$-even and $C$-odd multiplicities.}
\label{o4contd}
\end{table}

\begin{table}
\renewcommand{\arraystretch}{1.3}
\setlength{\tabcolsep}{4pt}
\centering
\begin{tabular}[t]{|c||c|c||c|c||c|c|}
\hline
    $\Wc$ & $\partial\,\Omega^\pm(1)$ & $\partial\,\Omega^\pm_{os}(1)$ & $\partial\,\Omega^\pm(2)$ & $\partial\,\Omega^\pm_{os}(2)$ & $\partial\,\Omega^\pm(3)$ & $\partial\,\Omega^\pm_{os}(3)$ \\
\hline
\hline
$\tau^{(1)}_1$ & 0, 1 & 0, 1 & 2, 1 & 1, 0 & 4, 6 & 2, 4 \\
\hline
$\tau^{(1)}_2$ & 0, 0 & 0, 0 & 0, 0 & 0, 0 & 1, 2 & 1, 2 \\
\hline
$\tau^{(1)}_3$ & 0, 0 & 0, 0 & 0, 0 & 0, 0 & 2, 1 & 2, 1 \\
\hline
$\tau^{(1)}_4$ & 1, 0 & 1, 0 & 2, 1 & 1, 0 & 6, 4 & 4, 2 \\
\hline
$\tau^{(2)}_1$ & 0, 0 & 0, 0 & 1, 0 & 1, 0 & 4, 5 & 3, 4 \\
\hline
$\tau^{(2)}_2$ & 0, 0 & 0, 0 & 1, 0 & 1, 0 & 5, 4 & 4, 3 \\
\hline
$\tau^{(3)}_1$ & 0, 1 & 0, 1 & 3, 2 & 2, 1 & 8, 11 & 4, 7 \\
\hline
$\tau^{(3)}_2$ & 0, 0 & 0, 0 & 1, 1 & 1, 1 & 6, 6 & 4, 4 \\
\hline
$\tau^{(3)}_3$ & 0, 0 & 0, 0 & 1, 1 & 1, 1 & 6, 6 & 4, 4 \\
\hline
$\tau^{(3)}_4$ & 1, 0 & 1, 0 & 3, 2 & 2, 1 & 11, 8 & 7, 4 \\
\hline
$\tau^{(4)}_1$ & 1, 1 & 1, 1 & 4, 4 & 2, 3 & 15, 14 & 8, 7 \\
\hline
$\tau^{(4)}_2$ & 0, 0 & 0, 0 & 1, 1 & 1, 1 & 8, 7 & 6, 5 \\
\hline
$\tau^{(4)}_3$ & 0, 0 & 0, 0 & 1, 1 & 1, 1 & 8, 7 & 6, 5 \\
\hline
$\tau^{(4)}_4$ & 1, 1 & 1, 1 & 4, 4 & 3, 2 & 15, 14 & 8, 7 \\
\hline
$\tau^{(6)}_1$ & 1, 1 & 1, 1 & 4, 5 & 2, 3 & 19, 19 & 11, 11 \\
\hline
$\tau^{(6)}_2$ & 0, 0 & 0, 0 & 2, 1 & 2, 1 & 12, 12 & 9, 9 \\
\hline
$\tau^{(6)}_3$ & 0, 1 & 0, 1 & 4, 3 & 3, 2 & 14, 17 & 8, 11 \\
\hline
$\tau^{(6)}_4$ & 1, 0 & 1, 0 & 4, 3 & 3, 2 & 17, 14 & 11, 8 \\
\hline
$\tau^{(8)}_1$ & 0, 1 & 0, 1 & 4, 4 & 3, 3 & 20, 20 & 13, 12 \\
\hline
$\tau^{(8)}_2$ & 0, 1 & 0, 1 & 4, 4 & 3, 3 & 20, 20 & 13, 12 \\
\hline
\end{tabular}
    \caption{Multiplicity of $\Wc$ representations in the total derivative
    spaces $\partial\,\Omega^\pm(k)$ and the on-shell $\partial\,\Omega^\pm_{os}(k)$. Each multiplicity is split into
    $C$-even and $C$-odd multiplicities.}
    \label{deltatau}
\end{table}

\begin{table}
\renewcommand{\arraystretch}{1.3}
\setlength{\tabcolsep}{4pt}
\centering
\begin{tabular}[t]{|c||c|c||c|c||c|c|}
\hline
    $\W$ & $\partial\,\Omega^\pm(1)$ & $\partial\,\Omega^\pm_{os}(1)$ & $\partial\,\Omega^\pm(2)$ & $\partial\,\Omega^\pm_{os}(2)$ & $\partial\,\Omega^\pm(3)$ & $\partial\,\Omega^\pm_{os}(3)$ \\
\hline
\hline
$\varrho^{(1)}_{1}$ & 0, 1 & 0, 1 & 2, 1 & 1, 0 & 3, 4 & 1, 2 \\
\hline
$\varrho^{(1)}_{2}$ & 0, 0 & 0, 0 & 0, 0 & 0, 0 & 0, 0 & 0, 0 \\
\hline
$\varrho^{(1)}_{3}$ & 0, 0 & 0, 0 & 0, 0 & 0, 0 & 0, 0 & 0, 0 \\
\hline
$\varrho^{(1)}_{4}$ & 1, 0 & 1, 0 & 2, 1 & 1, 0 & 4, 3 & 2, 1 \\
\hline
$\varrho^{(2)}_{1}$ & 0, 0 & 0, 0 & 0, 0 & 0, 0 & 1, 2 & 1, 2 \\
\hline
$\varrho^{(2)}_{2}$ & 0, 0 & 0, 0 & 0, 0 & 0, 0 & 2, 1 & 2, 1 \\
\hline
$\varrho^{(2)}_{3}$ & 0, 0 & 0, 0 & 0, 0 & 0, 0 & 1, 2 & 1, 2 \\
\hline
$\varrho^{(2)}_{4}$ & 0, 0 & 0, 0 & 0, 0 & 0, 0 & 2, 1 & 2, 1 \\
\hline
$\varrho^{(4)}_{1}$ & 1, 1 & 1, 1 & 3, 3 & 1, 2 & 9, 8 & 4, 3 \\
\hline
$\varrho^{(4)}_{2}$ & 0, 0 & 0, 0 & 0, 0 & 0, 0 & 2, 1 & 2, 1 \\
\hline
$\varrho^{(4)}_{3}$ & 0, 0 & 0, 0 & 0, 0 & 0, 0 & 2, 1 & 2, 1 \\
\hline
$\varrho^{(4)}_{4}$ & 1, 1 & 1, 1 & 3, 3 & 2, 1 & 9, 8 & 4, 3 \\
\hline
$\varrho^{(4)}_{5}$ & 0, 0 & 0, 0 & 1, 0 & 1, 0 & 3, 3 & 2, 2 \\
\hline
$\varrho^{(6)}_{1}$ & 1, 1 & 1, 1 & 3, 4 & 1, 2 & 10, 10 & 4, 4 \\
\hline
$\varrho^{(6)}_{2}$ & 0, 0 & 0, 0 & 1, 0 & 1, 0 & 3, 3 & 2, 2 \\
\hline
$\varrho^{(8)}_{1}$ & 0, 0 & 0, 0 & 1, 1 & 1, 1 & 6, 6 & 4, 4 \\
\hline
$\varrho^{(8)}_{2}$ & 0, 0 & 0, 0 & 1, 1 & 1, 1 & 6, 6 & 4, 4 \\
\hline
$\varrho^{(8)}_{3}$ & 0, 0 & 0, 0 & 1, 1 & 1, 1 & 6, 6 & 4, 4 \\
\hline
$\varrho^{(8)}_{4}$ & 0, 0 & 0, 0 & 1, 1 & 1, 1 & 6, 6 & 4, 4 \\
\hline
$\varrho^{(9)}_{1}$ & 0, 1 & 0, 1 & 3, 2 & 2, 1 & 8, 11 & 4, 7 \\
\hline
$\varrho^{(9)}_{2}$ & 0, 0 & 0, 0 & 1, 1 & 1, 1 & 6, 6 & 4, 4 \\
\hline
$\varrho^{(9)}_{3}$ & 0, 0 & 0, 0 & 1, 1 & 1, 1 & 6, 6 & 4, 4 \\
\hline
$\varrho^{(9)}_{4}$ & 1, 0 & 1, 0 & 3, 2 & 2, 1 & 11, 8 & 7, 4 \\
\hline
$\varrho^{(12)}_{1}$ & 0, 0 & 0, 0 & 1, 1 & 1, 1 & 9, 9 & 7, 7 \\
\hline
$\varrho^{(16)}_{1}$ & 0, 1 & 0, 1 & 3, 3 & 2, 2 & 14, 14 & 9, 8 \\
\hline
\end{tabular}
    \caption{Multiplicity of $\W$ representations in the total derivative
    spaces $\partial\,\Omega^\pm(k)$ and the on-shell $\partial\,\Omega^\pm_{os}(k)$. Each multiplicity is split into
    $C$-even and $C$-odd multiplicities.}
    \label{deltarho}
\end{table}

Turning to the actual off-forward matrix elements, 
with $p \neq p^\prime$, let us introduce $\Delta = p^\prime - p,\; q = \tfrac12
(p+p^\prime)$ and $t = \Delta^2$. The parametrization of the off-forward matrix elements of twist-2 operators is
then,
\bea
&&\langle p^\prime | {\cal O}_{\mu_1 \ldots \mu_n} | p \rangle = \\
&=&{\bar u}(p^\prime) \left[ 
\sum_{k=0}^{\left[\tfrac{n-1}{2}\right]} \left( 
A_{n,2k}(t) \gamma_{\mu_1} + B_{n,2k}(t) \frac{ i \sigma_{\mu_1
\nu} \Delta^\nu  } {2 m_N }
\right) \Delta_{\mu_2} \ldots \Delta_{\mu_{2k+1}} q_{\mu_{2k+2}} \ldots q_{\mu_n} + \right. \nn \\
&+&\left. \frac{C_n(t)}{m_N} \Delta_{\mu_1} \ldots
\Delta_{\mu_n} 
\right]_{H_n} u(p) \nn
\eea
where $C_n(t)$ is non-zero only for even $n$. The $k=0$ coefficients $A_{n,0}(t)$ reduce to the moments of parton
distribution functions at $p = p^\prime$, $A_{n,0}(0) = v_n = \langle x^{n-1} \rangle$, which we have already analyzed
in detail. The $n=2$ coefficients are the gravitational form factors. 
Lattice determinations of the generalized form factors on the cubic lattice include
\cite{LHPC:2010jcs,Alexandrou:2011nr,Bali:2018zgl,Alexandrou:2019ali}, for the gravitational form factors
of the pion and of hadrons of other spins see \cite{Pefkou:2021fni,Hackett:2023nkr}, and for the
resulting quark and gluon angular momentum \cite{Mathur:1999uf,Deka:2013zha,Engelhardt:2017miy,Wang:2021vqy}.
The helicity flip, or tensor, generalized form factors were studied in \cite{Gockeler:2005cj,Alexandrou:2022dtc}.
Alternatively, the $x$-dependence of generalized parton distributions may be accessed through non-local
operators \cite{Alexandrou:2020zbe,Lin:2020rxa,CSSMQCDSFUKQCD:2021lkf,Bhattacharya:2022aob}.
For
some cubic lattice representations the one-loop mixing with total derivative operators was computed in
\cite{Gockeler:2006nb}. 

Mixing with total derivative operators occurs in the continuum as well, but only logarithmically. 
As can
be seen from table \ref{o4contd}, $H_2$ does not appear in $\partial\,\Omega^+_{os}(1)$, so the analysis does not
change relative to forward matrix elements. On the other hand, $H_3$ and $H_4$ appear 
in $\partial\,\Omega^-_{os}(2)$ and $\partial\,\Omega^+_{os}(3)$ once, respectively. The corresponding total
derivative operators are,
\bea
\label{d34}
H_3&:& \qquad \left. \partial_\mu \partial_\nu \left( {\bar\psi} \gamma_\rho \psi \right) \right|_{H_3} \\
H_4&:& \qquad \left. \partial_\mu \partial_\nu \left( {\bar\psi} \gamma_\rho D_\sigma \psi \right) \right|_{H_4}. \nn
\eea
This means that for off-forward matrix elements with $n=3$ and $n=4$ one
additional operator is present and the mixing matrix becomes $2 \times 2$.

On the lattice, at $n=2$ there is also no change relative to the forward matrix elements, the two
preferred channels, $\tau^{(3)}_1$ and $\varrho^{(9)}_1$, do not appear in $\partial\,\Omega^+_{os}(1)$, so the
analysis from $p=p^\prime$ carries over verbatim.
Hence on both lattices one $Z$-factor determines all three generalized form factors
$A_{20}(t), B_{20}(t), C_2(t)$ completely.

At $n=3$, the preferred candidates were $\tau^{(8)}_1$ and $\varrho^{(8)}_3$. On the cubic lattice
$\tau^{(8)}_1$ can be found in $\partial\,\Omega^-_{os}(1)$ leading to a $1/a$ divergence from a total
derivative lattice operator and also three times in $\partial\,\Omega^-_{os}(2)$ enlarging the mixing matrix to
$5 \times 5$. On the other hand, on the 16-cell honeycomb the pattern is the same as in the continuum,
there is a single appearance of $\varrho^{(8)}_3$ in $\partial\,\Omega^-_{os}(2)$, leading to a $2 \times
2$ mixing matrix. Hence on the 16-cell honeycomb all generalized form factors 
$A_{30}(t), B_{30}(t), A_{32}(t), B_{32}(t)$ can be determined without any power divergences.

At $n=4$, on the cubic lattice $\tau^{(6)}_1$ was the preferred channel, which appears in
$\partial\,\Omega^+_{os}(1)$ once, in $\partial\,\Omega^+_{os}(2)$ twice, and $\partial\,\Omega^+_{os}(3)$
eleven times. Hence new $1/a^2$ and $1/a$ divergences are present in off-forward matrix elements, and as
many as eleven new operators to mix with logarithmically.

The situation on the 16-cell honeycomb, for $n=4$, is again identical to the continuum, the preferred
channel from forward matrix elements was $\varrho^{(2)}_3$ and this only appears in
$\partial\,\Omega^+_{os}(3)$ once, just as in the continuum. Hence again all $A_{40}(t), B_{40}(t), C_4(t),
A_{42}(t), B_{42}(t)$ coefficients are completely fixed by a $2 \times 2$ logarithmic mixing matrix
without any power divergences.

In section \ref{fs2} we have discussed the flavor singlet and gluonic sector, which may also contribute
off-forward for $n=4$ and of course not for $n=2$ as no total derivative gluonic operator exists at mass
dimension 4. 
In the continuum, there is a single total derivative gluonic operator, given in
terms of the gluonic part of the traceless energy-momentum tensor $T^g_{\mu\nu} = 
\tr F_{\mu \alpha} F_{\nu \alpha} - \tfrac14 \delta_{\mu \nu} \tr F_{\alpha\beta}F_{\alpha\beta}$ as,
\bea
\left. \partial_{\mu_1} \partial_{\mu_2} T^g_{\mu_3 \mu_4} \right|_{H_4}.
\eea
For $n=4$ only the 16-cell honeycomb is relevant in practice, in which case there is a single lattice operator
corresponding to the above total derivative gluonic operator, hence the renormalization structure is
exactly as in the continuum, even in the off-forward flavor singlet channel.

In summary, on the 16-cell honeycomb, both forward and off-forward matrix elements of $n=2,3,4$ twist-2 operators
can be renormalized exactly as in the continuum, and all forward matrix elements can be
calculated with a nucleon at rest, a major improvement over the cubic lattice.

\subsection{Explicit 16-cell honeycomb operators in the rest frame}
\label{expl2}

In this section we collect all three twist-2 operators for $v_n = \langle x^{n-1}\rangle$ for $n=2,3,4$ which
are accessible on the 16-cell honeycomb in the rest frame for easy reference. 
What is required is that the kinematic factor
in (\ref{kin}) is non-zero for $p=(0,0,0,E)$. The corresponding operators are normalized such that we
have
\bea
\langle p | {\cal O}_{(n)} | p \rangle = \langle x^{n-1} \rangle E^n.
\eea
The three irreducible representations are $\varrho^{(9)}_1, \varrho^{(8)}_3, \varrho^{(2)}_3$, for $n=2,3,4$,
respectively, and we have,
\bea
{\cal O}_{(2)} &=& \frac23\, {\cal O}_{44} \nn \\
{\cal O}_{(3)} &=& {\cal O}_{444} \\
{\cal O}_{(4)} &=& {\cal O}_{1111} + {\cal O}_{2222} + {\cal O}_{3333} + {\cal O}_{4444}, \nn
\eea
using the original $H_n$-projected definition (\ref{twist2}) for ${\cal O}_{\mu_1\ldots\mu_n}$. 
For convenience, we also list below the combinations
{\em before} the projection to $H_n$, in terms of 
\bea
O_{\mu_1 \ldots \mu_n} = {\bar \psi} \gamma_{\mu_1} i D_{\mu_2} \ldots i D_{\mu_n} \psi,
\eea
so without any symmetrization or trace subtraction,
\bea
\label{vnexplicit}
O_{(2)} &=& \frac23 O_{44} - \frac16\sum_\mu O_{\mu\mu} \nn \\
O_{(3)} &=&  O_{444} - \frac16\sum_\mu\left(O_{4\mu\mu}+O_{\mu4\mu}+O_{\mu\mu4}\right) \nn \\
O_{(4)} &=& \frac12\sum_\mu O_{\mu\mu\mu\mu} - \frac16\sum_{\mu\neq\nu}\left(O_{\mu\mu\nu\nu}+O_{\mu\nu\mu\nu}+O_{\mu\nu\nu\mu}\right).
\eea

\section{Polarized twist-2 matrix elements and moments of helicity PDFs}
\label{twist2polsec}

All the machinery from the unpolarized case carries over to polarized matrix elements too. The
helicity counterpart of (\ref{twist2}) is the operator
\bea
\label{twist2pol}
\widetilde{\cal O}_{\mu_1\ldots\mu_n} = 
\left. \bar\psi \gamma_{\mu_1} \gamma_5 iD_{\mu_2}\ldots iD_{\mu_n}\psi \right|_{H_n},
\eea
which are in the $O(4)$ representation $\det \otimes H_n$. The reduced matrix elements $a_n$ are defined by
\bea
\langle p,S| \widetilde{\cal O}_{\mu_1\ldots\mu_n}|p,S\rangle = 
a_{n-1} \; \big. S_{\mu_1}p_{\mu_2}\ldots p_{\mu_n} \big|_{H_n}.
\eea

We have $C = (-1)^{n+1}$ so the charge conjugation selection rule is the opposite of section
\ref{twist2sec}, otherwise the discussion proceeds analogously. The mass dimension is $n+2$ and the
appearances of $\det\otimes H_n$ in $\Omega^{(-1)^{n+1}}_{os}(k)$ with $k \leq n - 1$, 
as can be seen from table \ref{o4cont}, 
is exactly one for
$k = n - 1$ and none otherwise, leading to the conclusion that in the continuum renormalization is
multiplicative for each operator on its own, without logarithmic mixing.

\subsection{Cubic lattice}

On the cubic lattice, the relevant $\det\otimes H_n$ representations decompose under $W(B_4)$ as given in
table \ref{o4decomp} and their multiplicity in $\Omega^{(-1)^{n+1}}_{os}(k)$ can be read off from table
\ref{omegatau}. Let us consider $n=1,2,3,4$ below.

For $a_0$ we have $\det\otimes H_1 = \tau^{(4)}_4$ and it renormalizes the same way as in the continuum
and it appears in (\ref{polcubic}) so it is usable at rest. Lattice determinations of $a_0$ and $a_1$ on
the cubic lattice include \cite{Lin:2008uz,Aoki:2010xg,Bali:2014nma,Gupta:2018qil,Mondal:2020ela,Rodekamp:2023wpe}.

For $a_1$ there are two components in $\det\otimes H_2$ and both renormalize the same way as in the
continuum, but only $\tau^{(6)}_4$ is usable at rest, while for $\tau^{(3)}_4$ at least one spatial
momentum component needs to be non-zero and the spin needs to be parallel with it. 

For $a_2$ we have three components in $\det\otimes H_3$ and their fate is,
\begin{itemize}
    \item $\tau^{(4)}_3$, renormalizes as in the continuum.
    \item $\tau^{(4)}_4$, appears in $\Omega^+(0)$ leading to a $1/a^2$ divergence and twice in
        $\Omega^+_{os}(2)$ enlarging the mixing to a $2\times 2$ matrix.
    \item $\tau^{(8)}_2$, appears in $\Omega^+_{os}(1)$ leading to a $1/a$ divergence and twice in
        $\Omega^+_{os}(2)$ enlarging the mixing to a $2\times 2$ matrix.
\end{itemize}
From (\ref{polcubic}) we see that only $\tau^{(4)}_4$ and $\tau^{(8)}_2$ are available at rest and
unfortunately these are exactly the ones which bring in power divergences. In $\tau^{(4)}_3$ at least one
spatial momentum component needs to be non-zero with an orthogonal spin. 

Finally for $a_3$ we have seven components in $\det\otimes H_4$ and their properties are,
\begin{itemize}
\item $\tau^{(1)}_3$, behaves exactly as in the continuum.
\item $\tau^{(1)}_4$, there are no power divergences, but it appears twice in 
    $\Omega^-_{os}(3)$, giving a $2\times2$ logarithmic mixing matrix.
\item $\tau^{(2)}_2$, no power divergence, but appears twice in 
    $\Omega^-_{os}(3)$ leading to a $2\times2$ logarithmic mixing matrix.
\item $\tau^{(3)}_4$, appears in $\Omega^-_{os}(1)$, leading to a $1/a^2$ divergence, 
    and four times in $\Omega^-_{os}(3)$ leading to a $4 \times 4$ mixing matrix.
\item $\tau^{(6)}_1$, appears in $\Omega^-(0)$, leading to a $1/a^3$ divergence,
    three times in $\Omega^-_{os}(2)$ with a $1/a$ divergence, and five times in $\Omega^-_{os}(3)$
    leading to a $5 \times 5$ mixing matrix.
\item $\tau^{(6)}_2$, appears once in $\Omega^-_{os}(2)$,
    leading to a $1/a$ divergence, and four times in $\Omega^-_{os}(3)$ resulting in a $4\times 4$ mixing
    matrix.
\item $\tau^{(6)}_4$, appears in $\Omega^-_{os}(1)$  and in
$\Omega^-_{os}(2)$ leading to $1/a^2$ and $1/a$ divergences, respectively, and also six times 
        in $\Omega^-_{os}(3)$, so a $6\times 6$ mixing matrix.
\end{itemize}
The kinematics is again unfortunately such that the components which are free from power divergences are
not usable at rest. The component $\tau^{(1)}_3$ needs at least two non-zero spatial momentum components 
and non-parallel spin,
while $\tau^{(1)}_4$ and $\tau^{(2)}_2$ need one with a spin parallel to it. The components usable at
rest, $\tau^{(6)}_{1,4}$, i.e. those which do appear in (\ref{polcubic}), suffer from power divergences.

\subsection{16-cell honeycomb}

On the 16-cell honeycomb the situation is much improved as we will see. The corresponding analysis again follows 
directly from tables \ref{o4decomp} and \ref{omegarho}. Whether any given representation is usable at
rest is decided by its appearance in (\ref{pol16}).

For $a_0$ we have $\det\otimes H_1 = \varrho^{(4)}_4$ and it renormalizes the same way as in the continuum
and it appears in (\ref{pol16}) so it is usable at rest.

For $a_1$ we have $\det\otimes H_2 = \varrho^{(9)}_4$, a single component, which again renormalizes the
same as in the continuum and is also usable at rest. 

For $a_2$ we have two components in $\det\otimes H_3 = \varrho^{(8)}_2 \oplus \varrho^{(8)}_4$ 
and both renormalize the same as in the continuum
and both are usable at rest. Note that having both properties simultaneously was not possible on the cubic
lattice. The fact that we have two equally favorable channels is useful as a cross-check in practice. The
continuum limits should be the same but at $O(a^2)$ they of course differ.

There are four components in $\det\otimes H_4$, needed for $a_3$, and their properties are as follows,
\begin{itemize}
\item $\varrho^{(2)}_2$, behaves exactly as in the continuum.
\item $\varrho^{(2)}_4$, also behaves exactly as in the continuum.
\item $\varrho^{(9)}_4$, appears in $\Omega^-_{os}(1)$, leading to $1/a^2$ divergences and
    four times in $\Omega^-_{os}(3)$ giving rise to a $4\times 4$ mixing matrix.
\item $\varrho^{(12)}_1$, appears once in $\Omega^-_{os}(2)$, leading to a $1/a$
    divergence, and three times in $\Omega^-_{os}(3)$ so will require a $3 \times 3$ mixing matrix. 
\end{itemize}
Both channels $\varrho^{(2)}_{2,4}$, which renormalize the same as in the continuum, require at least one spatial
momentum component non-zero and parallel spin. The other two components are usable at rest, but since
the power divergence is milder and logarithmic mixing smaller in $\varrho^{(12)}_1$, this channel is
favored over $\varrho^{(9)}_4$.

In summary, the most pronounced advantage of the 16-cell honeycomb over the cubic lattice is $a_2$, which
gives $\int dx x^2 g_1(x) = \tfrac14 a_2$ and also determines 
$\int dx x^2 g_2(x) = \tfrac16 ( d_2 - a_2 )$ through the
Wandzura-Wilczek decomposition with $d_2$ the twist-3 component, see section \ref{twist3sec}. On the
cubic lattice continuum-like renormalization and rest frame kinematics were not simultaneously possible, while both are
possible on the 16-cell honeycomb.

\subsection{Flavor singlet and gluonic operators}
\label{singlettwist2pol}

Let us analyze the mixing pattern of gluonic operators with the flavor singlet variants of operators of
the type (\ref{twist2pol}), similarly to the previous section concerning $v_n$.

Even though there is no mass dimension 3 gauge invariant gluonic operator to mix with $\widetilde O_\mu$, 
relevant for $a_0 = 2 g_A$, the nucleon axial charge \cite{FlavourLatticeAveragingGroupFLAG:2024oxs}, 
the $U(1)_A$ anomaly does introduce anomalous renormalization. This is
present on both lattices of course. For the quark spin contributions to the proton spin on the cubic
lattice see \cite{Lin:2018obj,Alexandrou:2020sml}, and for the flavor decomposition of the corresponding
$x$-dependent distributions \cite{Alexandrou:2021oih}. The gluon spin, which has no local gauge invariant
operator, was studied in \cite{Yang:2016plb}.

Returning to gauge invariant gluonic operators, 
in the continuum only $a_2$ is affected, due to the matching $C=1$ quantum number of the mass dimension 5 operator,
\bea
\label{gluonicdeth}
\widetilde{\cal O}^g_{\mu_1\mu_2\mu_3} = \big. \tr {\tilde F}_{\nu \mu_1} D_{\mu_2} F_{\nu \mu_3} \big|_{H_3},
\eea
which belongs to $\det \otimes H_3$. This enlarges the continuum mixing matrix to $2 \times 2$. 
On both lattices each component mixes with the corresponding
component of (\ref{gluonicdeth}) hence the continuum $2\times 2$ mixing matrix stays $2\times 2$ on the
cubic lattice with $\tau^{(4)}_3$ but becomes $3\times 3$ with $\tau^{(8)}_2$. On the 16-cell honeycomb,
both $\varrho^{(8)}_2$ and $\varrho^{(8)}_4$ behaved as in the continuum in the flavor non-singlet case,
hence for both mixing in the singlet sector will be $2 \times 2$, again the same as in the continuum.

In the continuum there is no gluonic mass dimension 6 operator to mix, as far as $a_3$ is concerned. The 
3-gluon operator with $C=-1$,
\bea
{\cal O}^d_{\nu\mu_1\mu_2\mu_3} = d_{abc} F^a_{\nu\alpha} F^b_{\alpha\mu_1} F^c_{\mu_2\mu_3},
\label{dg}
\eea
given in terms of the totally symmetric $d_{abc}$ tensor of the gauge group is dimension 6 but is
excluded. This is because it lies in $Sym^3(A)$ given by the totally symmetric combinations of the three
anti-symmetric field strength tensors and we have,
\bea
Sym^3(A) = \left[ (3,0) \oplus (0,3) \right] \oplus 2A \oplus M_4
\label{sym3a}
\eea
which does not contain $\det \otimes H_4$, the relevant representation for $a_3$. On both lattices,
however, mixing occurs if the decomposition of (\ref{sym3a}) shares a factor with the decomposition of
$\det\otimes H_4$. From table \ref{o4decomp} we can find the decomposition of $A$ and $M_4$ as well as
$\det\otimes H_4$ with respect
to both lattice groups and we also have,
\bea
(3,0) \oplus (0,3) &=& \tau^{(1)}_2 \oplus \tau^{(1)}_3 \oplus \tau^{(6)}_1 \oplus \tau^{(6)}_2 \nn \\
                   &=& \varrho^{(1)}_2 \oplus \varrho^{(1)}_3 \oplus \varrho^{(6)}_1 \oplus \varrho^{(6)}_2,
\eea
hence we see that the factors $\tau^{(1)}_3, \tau^{(6)}_1, \tau^{(6)}_2, \tau^{(6)}_4$ on the cubic
lattice and $\varrho^{(12)}_1$ appear in both $Sym^3(A)$ and $\det\otimes H_4$. In these channels, the
operator (\ref{dg}) mixes logarithmically. On the cubic lattice, $\tau^{(1)}_3$ had continuum-like
renormalization but two spatial momentum components needed to be non-zero and the other three had power
divergences. On the 16-cell honeycomb, the channels with continuum-like renormalization are different
from $\varrho^{(12)}_1$ and they required one non-zero spatial momentum component only, hence are not
affected. The channel $\varrho^{(12)}_1$ had a $1/a$ power divergence in the flavor non-singlet case and 
a $3 \times 3$ mixing matrix,
hence the only change in the flavor singlet case is that the logarithmic mixing is enlarged to $4 \times 4$.

\subsection{Generalized form factors}

Just as in section \ref{gff}, in the polarized case also generalized form factors with $p \neq p^\prime$
can have contributions from derivative operators. All the machinery set up so far can trivially be
extended. Let us parametrize the off-forward matrix elements as, 
\bea
&&\langle p^\prime, S^\prime | \widetilde {\cal O}_{\mu_1 \ldots \mu_n} | p, S \rangle = \\
&=&{\bar u}(p^\prime,s^\prime) \left[ 
\sum_{k=0}^{\left[\tfrac{n-1}{2}\right]} 
\left( 
\widetilde A_{n,2k}(t) \gamma_{\mu_1} + \widetilde B_{n,2k}(t) \frac{ \Delta_{\mu_1}} {2 m_N }
\right) 
\Delta_{\mu_2} \ldots \Delta_{\mu_{2k+1}} q_{\mu_{2k+2}} \ldots q_{\mu_n} \right]_{H_n} \hspace{-0.5cm} \gamma_5 \, u(p,s), \nn
\eea
where the spin label in $u(p,s)$ corresponds to the spin 4-vector $S$ via
$\bar u(p,s) \gamma_\mu \gamma_5 u(p,s) = 2 S_\mu$. The forward case is then $a_{n-1} = 2 \widetilde
A_{n,0}(0)$.

In the continuum $\det \otimes H_2$ does not appear in $\partial\Omega^-_{os}(1)$ as can be seen in table
\ref{o4contd} so there is nothing new off-forward relative to the forward case for $a_1$. However
$\det\otimes H_3$ and $\det\otimes H_4$ both appear once in $\partial\Omega^+_{os}(2)$ and
$\partial\Omega^-_{os}(3)$, respectively, and the total derivative operators are,
\bea
\label{dferm}
\det\otimes H_3 &:& \partial_{\mu_1} \partial_{\mu_2} \left.\left(\bar\psi \gamma_{\mu_3}\gamma_5\psi\right)\right|_{H_3} \\
\det\otimes H_4 &:& \partial_{\mu_1} \partial_{\mu_2} \left.\left(\bar\psi \gamma_{\mu_3}\gamma_5 D_{\mu_4}\psi\right)\right|_{H_4} \nn 
\eea
exactly paralleling (\ref{d34}) with a $\gamma_5$-insertion. The mixing matrix becomes $2 \times 2$.

On both lattices the $n=1$ and $n=2$ form factors follow the continuum. Since on the cubic lattice $n=3$
and $n=4$ were impossible with power divergence free renormalization at rest, let us focus on the 16-cell
honeycomb only. At $n=3$ we had two channels, $\varrho^{(8)}_{2,4}$, with continuum-like renormalization
and available at rest and both appear once in $\partial\Omega^+_{os}(2)$, again the same counting as
in the continuum. Simultaneously having a rest frame and only continuum-like renormalization was not
possible even on the 16-cell honeycomb at $n=4$ in the forward case. There were four components and the
doublets $\varrho^{(2)}_{2,4}$ renormalized continuum-like and this stays the same off-forward too, but
these required at least one non-zero spatial momentum component, while the remaining two were available
at rest but off-forward both receive $1/a$ power divergences and enlarged logarithmic mixing. 

In summary, the advantageous properties of the $n=3$ 16-cell operators, relevant for $a_2$, 
are maintained off-forward as well. On the cubic lattice the polarized generalized form factors were
computed with local operators in \cite{LHPC:2007blg,Bali:2018zgl,Alexandrou:2019ali}, and Mellin moments
of the axial-vector generalized parton distribution up to fifth order from non-local operators in
\cite{Bhattacharya:2024wtg}.

In section \ref{singlettwist2pol} we have discussed mixing among the flavor singlet and gluonic operators
in the forward case. In principle there could be gluonic total derivative operators which would mix
off-forward; however, it is easy to see that no such operator exists with the necessary mass dimension,
parity and charge conjugation quantum numbers. Hence in the continuum the logarithmic mixing matrix in
the flavor singlet sector off-forward stays $3\times 3$, the original fermionic and gluonic operators and
one fermionic total derivative (\ref{dferm}).

\subsection{Explicit 16-cell honeycomb operators}

The operators corresponding to channels available at rest with polarization in the $z$-direction are
collected below for future reference. Thus $p=(0,0,0,E)$ and $S=(0,0,E,0)$, and the operators are
normalized such that,
\bea
\langle p, S | \widetilde{\cal O}_{(n)} | p, S \rangle = a_{n-1} E^n.
\eea
For $a_1$ we have the $\varrho^{(9)}_4$ channel and for $a_2$ we have two, $\varrho^{(8)}_2$ and
$\varrho^{(8)}_4$, 
both equally usable in practice. The operators (\ref{twist2pol}) in the three cases explicitly are,
\bea
\widetilde {\cal O}_{(2)} &=& 2 \widetilde {\cal O}_{34} \\
\widetilde {\cal O}_{(3),1} &=& \frac92 \widetilde{\cal O}_{344} + \frac32 \widetilde{\cal O}_{333}, \nn \\
\widetilde {\cal O}_{(3),2} &=& -6\widetilde {\cal O}_{333} \nn 
\eea
so the projection onto $H_n$ was already done in the definition (\ref{twist2pol}). For convenience we
also list these below in terms of the unprojected operators,
\bea
\widetilde O_{\mu_1\ldots\mu_n} =  
\bar\psi \gamma_{\mu_1} \gamma_5 iD_{\mu_2}\ldots iD_{\mu_n}\psi,
\eea
so {\em before} any symmetrization or trace removal. The relevant combinations are,
\bea
\label{anexplicit}
\widetilde{\cal O}_{(2)} &=& \widetilde O_{34}+\widetilde O_{43} \\
\widetilde{\cal O}_{(3),1} &=& 
\widetilde O_{344} + \widetilde O_{434} + \widetilde O_{443}
-\frac12\widetilde O_{113}
-\frac12\widetilde O_{131}
-\frac12\widetilde O_{311}
-\frac12\widetilde O_{223}
-\frac12\widetilde O_{232}
-\frac12\widetilde O_{322} \nn \\
\widetilde{\cal O}_{(3),2} &=& 
\widetilde O_{113}+\widetilde O_{223} +\widetilde O_{443} +
\widetilde O_{131}+\widetilde O_{232} +\widetilde O_{434} +
\widetilde O_{311}+\widetilde O_{322} +\widetilde O_{344} - 3\widetilde O_{333}.  \nn 
\eea

\section{Polarized twist-3 matrix elements and moments of $g_2$}
\label{twist3sec}

The spin dependent structure functions, $g_1$ and $g_2$, also behave markedly differently on the cubic
and 16-cell honeycomb lattices. The pattern follows our discussion in the previous section, namely that
certain $O(4)$ representations decompose into fewer $\W$ irreducibles than into $\Wc$ irreducibles. 

Even moments of ${\bar g}_2(x)$ \cite{Shuryak:1981pi,Jaffe:1990qh} are determined by the matrix elements of the operators,
\bea
{\cal O}^5_{\nu\mu_1\ldots\mu_n} = \left. {\bar\psi} \gamma_\nu \gamma_5  i D_{\mu_1}
i D_{\mu_2} \ldots i D_{\mu_n} \psi\right|_{M_{n+1}},
\label{o5}
\eea
where the projection onto $M_{n+1}$, that is Young tableau $[n,1]$, is a projection onto 
$n+1$ indices of mixed $[n,1]$-type symmetry.
The mass dimension of ${\cal O}^5$ is $n+3$. Only the flavor non-singlet case is considered here; for the
flavor singlet and gluonic sectors see section \ref{fs3}. The
twist-3 reduced matrix elements $d_n$ are defined by,
\bea
\langle p, S | {\cal O}^5_{\nu\mu_1\ldots\mu_n} | p, S \rangle = d_n \;
\left.S_\nu p_{\mu_1} p_{\mu_2} \ldots p_{\mu_n}\right|_{M_{n+1}}.
\label{spp}
\eea
The corresponding representation, $\det\otimes M_{n+1} = M_{n+1}$, is irreducible under $O(4)$. The
projected operator is first made anti-symmetric in $(\nu,\mu_1)$ and then totally symmetric and traceless
in $(\mu_1,\ldots,\mu_n)$.

There are two main differences from twist-2 operators as far as renormalization is concerned.
First, there can be additional
operators of the same dimension, with which a logarithmic mixing can occur even in the continuum. 
We have $\det \otimes M_{n+1} = M_{n+1}$ and $C=(-1)^n$ and from table \ref{o4cont} we see explicitly that the
multiplicity of $M_3$ in $\Omega^+_{os}(2)$ is one, while the multiplicity of $M_4$ in $\Omega^-_{os}(3)$
is two. Generally, the multiplicity of $M_{n+1}$ in $\Omega^{(-1)^n}_{os}(n)$ is $n-1$. 
These $n-1$ operators can be constructed as follows, for $k=1,\ldots,n-1$,
\bea
R^{(k)}_{\nu\mu_1\ldots\mu_n} &=& \left.{\bar\psi} \;\, \gamma_{\mu_2}\;\; iD_{\mu_3} \ldots iD_{\mu_{k+1}} {\tilde
F_{\nu \mu_1}} iD_{\mu_{k+2}} \ldots iD_{\mu_n} \psi \right|_{M_{n+1}} \nn \\
S^{(k)}_{\nu\mu_1\ldots\mu_n} &=& \left.{\bar\psi} \gamma_{\mu_2} \gamma_5 iD_{\mu_3} \ldots iD_{\mu_{k+1}} 
F_{\nu \mu_1} iD_{\mu_{k+2}} \ldots iD_{\mu_n} \psi \right|_{M_{n+1}}
\label{rk}
\eea
where the label $k$ indicates where the field strength tensor or its dual is inserted in the chain of $n-2$ covariant
derivatives; the mass dimension is also $n+3$. Together these are $2(n-1)$ operators and charge
conjugation splits them equally into an $n-1$ dimensional even and also $n-1$ dimensional odd subspace.
Second, there is the operator,
\bea
{\cal O}^m_{\nu\mu_1\ldots\mu_n} =  \left. {\bar\psi} \sigma_{\nu\mu_1} \gamma_5 D_{\mu_2} \ldots
D_{\mu_n} \psi \right|_{M_{n+1}}
\label{om}
\eea
of mass dimension $n+2$, so $m \, {\cal O}^m$ can also mix.

The presence of the continuum operator ${\cal O}^m$ means that with Wilson fermions a $1/a$
subtraction needs to be done on both lattices. The different behavior between the cubic and the 16-cell
honeycomb will manifest itself in the different renormalization on top of this unavoidable $1/a$
divergence, which originates from (\ref{om}). 

We will only be interested in the matrix elements $d_2$ and $d_3$. At $n=2$ there is a single operator which
renormalizes multiplicatively, and ${\cal O}^5 = \tfrac14 R^{(1)}$ up to equations of motion and total
derivative terms. At $n=3$ we have a $2 \times 2$ logarithmic  mixing matrix, even in the continuum.

Up to the above two important differences relative to the twist-2 case, 
$1/a\,$ subtraction from the continuum operator (\ref{om}) 
and $(n-1) \times (n-1)$ logarithmic mixing
matrix as opposed to a single operator, the analysis on the lattice parallels the twist-2 discussion.
Additional lower dimensional lattice operators may generate power divergences and lattice operators of
the same dimension, $n+3$, may enlarge the $(n-1) \times (n-1)$ mixing matrix. 
On both lattices the key is the decomposition of $\det \otimes M_{n+1}$ with respect to $\Wc$ and $\W$. 
The enlargement of the $(n-1) \times (n-1)$ mixing matrix will occur
if an irreducible representation of the lattice symmetry group appears in both $\det \otimes M_{n+1}$
and $\det \otimes H_{n+1}$, because in that case a $\gamma_5$-inserted twist-2 lattice operator 
will additionally mix with the $n-1$ available twist-3 operators.
As a reminder, $M_{n+1} =(\tfrac{n+1}{2},\tfrac{n-1}{2}) \oplus (\tfrac{n-1}{2},\tfrac{n+1}{2})$ 
and so its dimension is $2n(n+2)$. 

The explicit projected momentum-spin 
tensor structure in (\ref{spp}) for $n=2$ in the continuum in Euclidean signature is,
\bea
t_{\nu\mu_1\mu_2}&=&
\frac13\left(S_\nu p_{\mu_1}-S_{\mu_1}p_\nu\right)p_{\mu_2}
+\frac13\left(S_\nu p_{\mu_2}-S_{\mu_2}p_\nu\right)p_{\mu_1} +
\nn \\
&+&\frac{p^2}{9}\left(
\delta_{\nu\mu_1}S_{\mu_2}+\delta_{\nu\mu_2}S_{\mu_1}-2\delta_{\mu_1\mu_2}S_\nu
\right),
\label{t2}
\eea
while for $n=3$,
\bea
\label{t3}
&& \qquad\qquad t_{\nu\mu_1\mu_2\mu_3}= \\
&=& \frac14\Big[(S_\nu p_{\mu_1}-S_{\mu_1}p_\nu)p_{\mu_2}p_{\mu_3}
    +(S_\nu p_{\mu_2}-S_{\mu_2}p_\nu)p_{\mu_1}p_{\mu_3}
    +(S_\nu p_{\mu_3}-S_{\mu_3}p_\nu)p_{\mu_1}p_{\mu_2}\Big] + \nn \\
&+&\frac{1}{16}p^2\Big[\delta_{\nu\mu_1}(S_{\mu_2}p_{\mu_3}+S_{\mu_3}p_{\mu_2})
+\delta_{\nu\mu_2}(S_{\mu_1}p_{\mu_3}+S_{\mu_3}p_{\mu_1})
+\delta_{\nu\mu_3}(S_{\mu_1}p_{\mu_2}+S_{\mu_2}p_{\mu_1})\Big] + \nn \\
&+&\frac{1}{48}p^2\Big[\delta_{\mu_1\mu_2}(S_{\mu_3}p_\nu-7S_\nu p_{\mu_3})
+\delta_{\mu_1\mu_3}(S_{\mu_2}p_\nu-7S_\nu p_{\mu_2})
+\delta_{\mu_2\mu_3}(S_{\mu_1}p_\nu-7S_\nu p_{\mu_1})\Big] \nn
\eea
and these will be further projected onto the irreducible representations of the lattice symmetry groups. 
Deciding whether polarized nucleons at rest, the most favorable kinematics on the lattice, 
can be used for the measurement of the matrix elements $d_{2,3}$, will be based on the discussion in
section \ref{restframe}.

The details of the cubic lattice setup are well-known for $d_2$
\cite{Gockeler:1995wg,Gockeler:2000ja,Gockeler:2005vw}, measured experimentally in
\cite{E155:2002iec,JeffersonLabHallA:2014gzr,SANE:2018pwx}, and recently extracted on the lattice also
from non-local operators \cite{Gao:2026wlz}, but there appears to be no full group
theoretic, kinematical and renormalization analysis for $d_3$. 
In any case the details will be given for both the cubic and the 16-cell honeycomb and both $d_2$ and
$d_3$ below for consistency. 

\subsection{Cubic lattice}

As far as $d_2$ is concerned,
from table \ref{o4decomp} we see that $M_3 = \tau^{(8)}_1 \oplus \tau^{(8)}_2$ and from (\ref{redund})
and (\ref{taudef}) it follows that $\det$ interchanges the two components.
The mass dimension is 5. Lower
dimensional operators may come from $\Omega^+(0)$ or $\Omega^+_{os}(1)$, 
corresponding to mass dimensions 3 and 4,
respectively. From table \ref{omegatau} we see that 
neither $\tau^{(8)}_1$ nor $\tau^{(8)}_2$ appears in $\Omega^+(0)$ and both appear once in $\Omega^+_{os}(1)$
which is just the lattice occurrence of the aforementioned mass term, also there in the continuum. 
Same dimension operators may come from $\Omega^+_{os}(2)$, and $\tau^{(8)}_1$ appears once,
$\tau^{(8)}_2$ appears twice. Hence the $\tau^{(8)}_1$ 
channel fully behaves like the continuum and the $\tau^{(8)}_2$ channel requires a $2 \times 2$
logarithmic mixing matrix when in the continuum we had a single operator.

The concrete decomposition of the tensor (\ref{t2}) under $\Wc$ is that
repeated indices span $\tau^{(8)}_2$ and all different indices span $\tau^{(8)}_1$. For a polarized
nucleon in the $z$-direction and at rest, we have,
\bea
t_{344}=\frac49E^3,\quad t_{434}=t_{443}=t_{311}=t_{322}=-\frac29E^3,\quad
t_{113}=t_{131}=t_{223}=t_{232}=\frac19E^3 ,
\label{ttt}
\eea
all non-zero, lying in $\tau^{(8)}_2$, which does appear in (\ref{polcubic}). The other,
$\tau^{(8)}_1$ is only non-vanishing if at least one spatial momentum component is non-zero
with orthogonal spin.
Unfortunately, continuum-like renormalization in the rest frame cannot occur simultaneously, 
only either one or the other practical property is possible to retain, although no power divergences are
present in either case. 

The corresponding analysis for $d_3$, mass dimension 6, $C=-1$, starts with the six components of $\det\otimes
M_4 = M_4$ with respect to $\Wc$, as can be seen from table \ref{o4decomp}. Here $\det$ interchanges
$\tau^{(3)}_{2,3}$ as well as $\tau^{(6)}_{3,4}$, while it leaves $\tau^{(6)}_{1,2}$ fixed.
Their occurrence in $\Omega^-(0)$ and $\Omega^-_{os}(1)$, potentially more than one occurrence in
$\Omega^-_{os}(2)$ and more than two occurrences in $\Omega^-_{os}(3)$
can be read off from table \ref{omegatau}. The fate of all six components is listed below,
\begin{itemize}
    \item $\tau^{(3)}_2$, $\tau^{(3)}_3$, both behave the same as in the continuum. 
    \item $\tau^{(6)}_1$, appears in $\Omega^-(0)$, leading to a $1/a^3$ divergence and also three times
        in $\Omega^-_{os}(2)$ and five times in $\Omega^-_{os}(3)$ enlarging the logarithmic mixing
        matrix to $5\times 5$.
    \item $\tau^{(6)}_2$, appears four times in $\Omega^-_{os}(3)$, hence free from extra power
        divergences but the logarithmic mixing is enlarged to $4 \times 4$. 
    \item $\tau^{(6)}_3$, appears three times in $\Omega^-_{os}(3)$, hence also free from extra power
        divergences and the logarithmic mixing matrix is $3 \times 3$. 
    \item $\tau^{(6)}_4$, appears in $\Omega^-_{os}(1)$ leading to a $1/a^2$ divergence, and also appears
        six times in $\Omega^-_{os}(3)$.
\end{itemize}
Kinematically, none of the above channels without power divergences are favorable, all polynomials vanish in the rest
frame. The channels which did not introduce additional power divergences require the following spin and
momentum configurations: $\tau^{(3)}_2$ at least 1 non-zero spatial momentum component and spin needs to
be parallel, $\tau^{(3)}_3$ at least 2 non-zero spatial momentum components and non-parallel spin, 
$\tau^{(6)}_{2,3}$ at least one
non-zero spatial momentum component and the spin needs to be orthogonal.

In summary, the determination of $d_3$ on a cubic lattice cannot be performed in the rest frame and
the most promising channels are $\tau^{(3)}_2$ where renormalization is continuum-like but is only
available for non-zero momenta and parallel spin, and $\tau^{(6)}_3$ where spin is orthogonal but spatial
momenta must have at least 1 non-zero component and renormalization is only complicated by a larger
logarithmic mixing matrix.

\subsection{16-cell honeycomb}

The $O(4)$ representation $M_3$ is irreducible with respect to $\W$, $M_3 = \det \otimes M_3 
= \varrho^{(16)}_1$, the largest irreducible representation, see table \ref{o4decomp}. This means the 
renormalization properties of $d_2$ on the 16-cell honeycomb 
are the same as in the continuum, in particular $m {\cal O}^m$ from (\ref{om}) mixes logarithmically
and $\varrho^{(16)}_1$ appears only once in both $\Omega^+_{os}(1)$ and $\Omega^+_{os}(2)$.
Since $\varrho^{(16)}_1 = \tau^{(8)}_1 \oplus
\tau^{(8)}_2$ under $\Wc$, all 16 corresponding cubic lattice polynomials from the previous section
are in the same multiplet, hence the non-zero components
(\ref{ttt}) can be used for a polarized nucleon in the
$z$-direction and at rest. This is consistent with $\varrho^{(16)}_1$ appearing in (\ref{pol16}).

Now for the matrix element $d_3$ we have three $\W$ components, 
$M_4 = \varrho^{(9)}_2 \oplus \varrho^{(9)}_3 \oplus \varrho^{(12)}_1$ from table \ref{o4decomp} 
and $\det$ interchanges the two 9 dimensional components while leaves $\varrho^{(12)}_1$ fixed.
From table \ref{omegarho} we also see that none of them appear in $\Omega^-(0)$ or $\Omega^-_{os}(1)$ and 
all three appear once in $\Omega^-_{os}(2)$, which are all the lattice versions of the continuum 
${\cal O}^m$ operator in the given representation for $n=3$. The occurrence in $\Omega^-_{os}(3)$
varies by channel and is listed below,
\begin{itemize}
    \item $\varrho^{(9)}_2$, behaves exactly as in the continuum.
    \item $\varrho^{(9)}_3$, also behaves exactly as in the continuum. 
    \item $\varrho^{(12)}_1$, the only difference from the continuum is that it appears three 
        times in $\Omega^-_{os}(3)$ enlarging the mixing matrix to $3 \times 3$.
\end{itemize}
Hence all 3 channels are free from new power divergences. The kinematics are the following,
$\varrho^{(9)}_2$ is not possible at rest, at least one spatial momentum component needs to be non-zero,
while both $\varrho^{(9)}_3$ and $\varrho^{(12)}_1$ are possible at rest, they both appear in
(\ref{pol16}). The independent non-zero tensor components of (\ref{t3}) are,
\bea
\varrho^{(9)}_3&:& t_{3444}=t_{4333}=\frac{3}{16} E^4,\qquad t_{1134}=t_{2234}=\frac{1}{16}E^4 \nn \\
&&t_{3114}=t_{3224}=t_{3334}=t_{4113}=t_{4223}=t_{4344}=-\frac{1}{16}E^4 \\
\varrho^{(12)}_1&:& t_{3444}=-t_{4333}=\frac{1}{8}E^4,\qquad t_{3334}=-t_{4344}=\frac{1}{24}E^4 \nn\\
&&t_{3114}=t_{3224}=-t_{4113}=-t_{4223}=-\frac{1}{12}E^4 \nn
\eea
Overall, the determination of both $d_2$ and $d_3$ is without issues on the 16-cell honeycomb using the
channels $\varrho^{(16)}_1$ and  $\varrho^{(9)}_3$, respectively; 
the renormalization patterns are the same as in the continuum and a nucleon at rest can be
used, something which was not simultaneously possible on the cubic lattice.

\subsection{Flavor singlet and gluonic operators}
\label{fs3}

Along the lines of section \ref{fs2}, where the twist-2 flavor singlet sector was analyzed, in this
section we do the same briefly for the twist-3 sector.

In the continuum there is no new independent gluonic operator at $n=2$ at all, while there is a single one
at $n=3$, encountered already in (\ref{dg}).
Hence it enlarges the logarithmic mixing for $d_3$ by one because as we have seen $Sym^3(A)$ contains
$M_4$ once.

On both lattices, at both $n=2$ and $n=3$, no new power divergences appear. On the cubic lattice,
there were two channels, $\tau^{(8)}_{1,2}$ and since $\det \otimes H_3 = \tau^{(4)}_3 \oplus
\tau^{(4)}_4 \oplus \tau^{(8)}_2$, we see that the latter does appear. The corresponding 
mass dimension 5, $C=1$ operator is
$\tr {\tilde F}_{\alpha\mu_1} D_{\mu_2} F_{\alpha \mu_3}$, projected onto $\tau^{(8)}_2$. This argument
is similar to the fermionic case in that new operators appear if $\det\otimes H_{n+1}$ and $\det\otimes M_{n+1}$
share a factor. Hence there is no additional gluonic mixing in the $\tau^{(8)}_1$ channel, 
but in the $\tau^{(8)}_2$ channel the mixing matrix becomes $3 \times 3$.

On the 16-cell honeycomb, on the other hand, both $n=2$ and $n=3$ behave exactly as in the continuum,
just the operator (\ref{dg}) appears and nothing else. Hence the good renormalization properties in the
flavor non-singlet sector are not spoiled by the flavor singlet and gluonic operators.

\subsection{Off-forward matrix elements}

Twist-3 generalized parton distributions were recently studied on the lattice using non-local operators
\cite{Bhattacharya:2023nmv}; here we consider the local operators which determine their moments.
Just as in section \ref{gff}, renormalization of off-forward matrix elements with $p \neq p^\prime$ may
introduce total derivative operators, which could have been discarded at $p=p^\prime$. In the continuum,
for $d_2$, corresponding to $M_3$ and mass dimension 5, we see from table \ref{o4contd} that no new power
divergences will occur since $M_3$ does not appear in $\partial\,\Omega^+_{os}(1)$. However, it does appear in
$\partial\,\Omega^+_{os}(2)$ two times. The corresponding operators are,
\bea
\left. \partial_{\mu_1} \left( \bar\psi \gamma_\nu D_{\mu_2} \psi \right)\right|_{M_3} \qquad \qquad 
\left. \partial_{\mu_1} \partial_{\mu_2} \left( \bar\psi \gamma_\nu \gamma_5 \psi \right)\right|_{M_3},
\eea
which leads to a $3 \times 3$ mixing matrix as opposed to a single multiplicatively renormalized
operator. For $d_3$ the relevant representation is $M_4$, mass dimension 6 and again from table
\ref{o4contd} we see that there is a new $1/a$ power divergent operator because $M_4$ appears in
$\partial\,\Omega^-_{os}(2)$ once, and four new logarithmically mixing operators because $M_4$ appears in
$\partial\,\Omega^-_{os}(3)$ four times. Specifically, we have
\bea
\partial\,\Omega^-_{os}(2)&:& 
\left.\partial_{\mu_2} \partial_{\mu_3} \left( \bar\psi \sigma_{\nu\mu_1} \psi \right)\right|_{M_4} \\
\partial\,\Omega^-_{os}(3)&:& 
\left. \partial_{\mu_3} \left( \bar\psi \gamma_{\mu_1} \gamma_5 F_{\nu \mu_2} \psi \right)\right|_{M_4}, \qquad
\left. \partial_{\mu_1} \left( \bar\psi \gamma_\nu D_{\mu_2} D_{\mu_3} \psi \right)\right|_{M_4} \nn \\
&&
\left.\partial_{\mu_2} \partial_{\mu_3} \left( \bar\psi \gamma_{\nu} \gamma_5 D_{\mu_1} \psi \right)\right|_{M_4},
\qquad
\left.\partial_{\mu_1} \partial_{\mu_2} \partial_{\mu_3} \left( \bar\psi \gamma_\nu \psi \right)\right|_{M_4} \nn 
\eea
All of this was in terms of continuum $O(4)$ irreducibles.

For $d_2$, on the cubic lattice, so mass dimension 5 and $C=1$, we had two components, $M_3 =
\tau^{(8)}_1 \oplus \tau^{(8)}_2$. Both appear three times in $\partial\,\Omega^+_{os}(2)$ and not below,
so there are no new power divergences but instead of the two continuum total derivative operators we have
three in both lattice channels. Hence in off-forward matrix elements the continuum $3\times 3$ mixing
matrix enlarges to $4\times 4$ in the channel $\tau^{(8)}_1$ and $5 \times 5$ in $\tau^{(8)}_2$. On the
16-cell honeycomb we had $\det \otimes M_3 = \varrho^{(16)}_1$ and this appears in
$\partial\,\Omega^+_{os}(2)$ exactly twice, and none below, the same numbers as in the continuum. Hence
the observed continuum-like renormalization of $d_2$ in forward matrix elements on the 16-cell honeycomb
persists in off-forward matrix elements as well.

As far as $d_3$ is concerned, on the cubic lattice, we had six components for $M_4$ and
their occurrence in $\partial\,\Omega^-_{os}(k)$ for $k=1,2,3$ is the same as in the continuum for
$\tau^{(3)}_2$ and $\tau^{(3)}_3$ only, while $\tau^{(6)}_2$ is also free from new power divergences but
the logarithmic mixing matrix enlarges from the forward $4 \times 4$ to $13 \times 13$ off-forward. In
any case, $d_3$ on the cubic lattice was unfavorable in forward matrix elements already.

On the other hand, on the 16-cell honeycomb we had three components, $M_4 = \varrho^{(9)}_2 \oplus
\varrho^{(9)}_3 \oplus \varrho^{(12)}_1$ and the occurrence of $\varrho^{(9)}_{2,3}$ in
$\partial\,\Omega^-_{os}(k)$ for $k=1,2,3$ is exactly the same as in the continuum and $\varrho^{(12)}_1$
appears only differently in $\partial\,\Omega^-_{os}(3)$, instead of 4 times, it appears 7 times. Hence,
the channel $\varrho^{(9)}_3$ which was kinematically favorable and had the same renormalization pattern
as in the continuum in forward matrix elements, retains all of these properties in off-forward matrix
elements as well.

The last aspect, similarly to the twist-2 case, is the question of gluonic total derivative operators
contributing to off-forward matrix elements in the flavor singlet channel. These only appear for $d_2$
and in the continuum the relevant space is $R \otimes Sym^2(A)$ where $R$ carries the derivative index
and $Sym^2(A)$ carries the pair of field strength tensors. We have the decomposition (\ref{rep21}) for
$Sym^2(A)$ and since $R \otimes H_2 = R \oplus H_3 \oplus M_3$ and also $R \otimes W =
(\tfrac52,\tfrac12) \oplus (\tfrac12,\tfrac52) \oplus M_3$, with $M_3$ appearing once in only these two
components, we are led to two total derivative gluonic operators with $C=1$ at mass dimension 5. These
are,
\bea
\left. \partial_{\mu_1} T^g_{\nu \mu_2} \right|_{M_3} 
\qquad \qquad \qquad \qquad \partial_\alpha W_{\alpha \nu \mu_1 \mu_2},
\label{totderg}
\eea
where $W_{\mu\nu\rho\sigma}$ is the Weyl-tensor part of $F F$, i.e. the $W$ component in (\ref{rep21}).  

On the cubic lattice we have the decompositions, from table \ref{o4decomp},  
$H_2 = \tau^{(3)}_1 \oplus \tau^{(6)}_3$ and $W = \tau^{(2)}_1 \oplus \tau^{(2)}_2 \oplus \tau^{(6)}_2$
and instead of a single irreducible $M_3$ we have $M_3 = \tau^{(8)}_1 \oplus \tau^{(8)}_2$, hence we need
to find the occurrence of both $\tau^{(8)}_{1,2}$ in the tensor products $R \otimes H_2$ and $R \otimes
W$. We have
\bea
\label{wcubic}
H_2:&& 
R \otimes \tau^{(3)}_1 = \tau^{(4)}_1 \oplus \tau^{(8)}_1, \qquad \qquad
R \otimes \tau^{(6)}_3 = \tau^{(4)}_1 \oplus \tau^{(4)}_2 \oplus \tau^{(8)}_1 \oplus \tau^{(8)}_2  \\
W:&&  
R \otimes \tau^{(2)}_1 = \tau^{(8)}_1, \qquad
R \otimes \tau^{(2)}_2 = \tau^{(8)}_2, \qquad
R \otimes \tau^{(6)}_2 = \tau^{(4)}_2 \oplus \tau^{(4)}_3 \oplus \tau^{(8)}_1 \oplus \tau^{(8)}_2. \nn
\eea
Clearly, $\tau^{(8)}_1$ appears four times, while $\tau^{(8)}_2$ appears three times, instead of the two 
in the continuum. Hence in the flavor
singlet off-forward channel the full continuum $5 \times 5$ mixing matrix becomes $8 \times 8$ and $9 \times
9$ in the channels $\tau^{(8)}_1$ and $\tau^{(8)}_2$, respectively, after taking into account 
the enlargements from forward gluonic, off-forward fermionic and the just discussed off-forward gluonic
operators.

On the 16-cell honeycomb the $\varrho^{(16)}_1$ channel behaved as in the continuum in forward matrix
elements, both flavor non-singlet and singlet, as well as off-forward flavor non-singlet channels. 
The only lattice artifact is in the off-forward flavor singlet channel, where the continuum $5\times 5$
mixing matrix is enlarged to $6 \times 6$, given by an additional total derivative gluonic operator,
beyond the two continuum ones (\ref{totderg}). This can be seen from the fact that $H_2 =
\varrho^{(9)}_1$ stays irreducible, and $W = \varrho^{(4)}_5 \oplus \varrho^{(6)}_2$ has only two
components, not three as on the cubic lattice. Furthermore $M_3 = \varrho^{(16)}_1$ is also irreducible.
The decompositions analogous to (\ref{wcubic}) are,
\bea
H_2:&& 
R \otimes \varrho^{(9)}_1 = \varrho^{(4)}_1 \oplus \varrho^{(8)}_1 \oplus \varrho^{(8)}_3 \oplus \varrho^{(16)}_1 \label{w16} \\
W:&&  
R \otimes \varrho^{(4)}_5 = \varrho^{(16)}_1, \qquad
R \otimes \varrho^{(6)}_2 = \varrho^{(4)}_2 \oplus \varrho^{(4)}_3 \oplus \varrho^{(16)}_1,  \nn
\eea
and $\varrho^{(16)}_1$ appears twice in the $W$ factor, not once as in the continuum, giving rise to an
additional operator on the 16-cell honeycomb.

In summary, twist-3 operators can be constructed on the 16-cell honeycomb which 
in all aspects behave as in the continuum for both $d_2$ and $d_3$,
forward and off-forward, flavor non-singlet and singlet channels, except for the off-forward singlet
case of $d_2$, where the only difference is one additional same mass dimension operator enlarging the logarithmic
mixing matrix size by one.

\subsection{Explicit 16-cell honeycomb operators in the rest frame}
\label{expl3}

Similarly to section \ref{expl2}, in this section we collect both twist-3 operators 
for $d_{2,3}$ which are accessible on the 16-cell honeycomb for a nucleon at rest and polarized in the
$z$-direction. The operators here are normalized such that we have,
\bea
\langle p, S | {\cal O}^5_{(n)} | p, S \rangle = d_n E^{n+1}.
\eea
The two irreducible representations are $\varrho^{(16)}_1, \varrho^{(9)}_3$, for $d_2, d_3$,
respectively, and we have,
\bea
{\cal O}^5_{(2)} &=& \frac94\, {\cal O}^5_{344} \\
{\cal O}^5_{(3)} &=& - 8 \left( {\cal O}^5_{3433} + {\cal O}^5_{4344} \right), \nn
\eea
using the original $M_{n+1}$-projected definition (\ref{o5}) for ${\cal O}^5_{\nu\mu_1\ldots\mu_n}$. 
For convenience, we also list below the combinations
{\em before} the projection to $M_{n+1}$, so in terms of 
\bea
O^5_{\nu\mu_1\ldots\mu_n} = 
{\bar\psi} \gamma_\nu \gamma_5  i D_{\mu_1} i D_{\mu_2} \ldots i D_{\mu_n} \psi 
\eea
without any symmetrization or anti-symmetrization or trace subtraction,
\bea
\label{dnexplicit}
O^5_{(2)}&=&O^5_{344}-O^5_{434}+\frac12\left(O^5_{131}+O^5_{232}\right)-\frac12\left(O^5_{311}+O^5_{322}\right) \\
O^5_{(3)}&=&O^5_{3444}-O^5_{3433}+O^5_{4333}-O^5_{4344} + \nn \\
&+&\frac12\left(O^5_{1314}+O^5_{1341}+O^5_{1413}+O^5_{1431}+O^5_{2324}+O^5_{2342}+O^5_{2423}+O^5_{2432}\right)- \nn \\
&-&\frac12\left(O^5_{3114}+O^5_{3141}+O^5_{3224}+O^5_{3242}+O^5_{4113}+O^5_{4131}+O^5_{4223}+O^5_{4232}\right). \nn
\eea

\section{Four-fermion operators}

So far we have considered quark bilinear operators, but at mass dimension 6, four-fermion operators may enter as
well. These may only affect $v_4 = \langle x^3 \rangle$, $a_3$ or $d_3$ since these have corresponding
operators of mass dimension 6. The relevant representations in the continuum are $H_4, \det\otimes H_4$
and $M_4$, respectively.

On dimensional grounds the four-fermion operators in question cannot contain derivatives,
and so must be of the form, schematically, $\left(\bar\psi \Gamma_1
\psi\right)\left(\bar\psi \Gamma_2 \psi\right)$, given by two $\gamma$-matrix structures
$(\Gamma_1,\Gamma_2)$. 
The color and flavor structure may also be non-trivial. As far as color is concerned, since
the operator should be a singlet, the two factors may be either $(1,1)$ or $({\rm adj},{\rm adj})$. The
flavor representation, on the other hand, can be non-singlet as well of course. More precisely, the flavor
structure can be one of $(1,1)$, $(1,{\rm adj})$, $({\rm adj},1)$, $({\rm adj},{\rm adj})$, where the first is
relevant for the flavor singlet case, the second and third only for the flavor non-singlet case, while the
last one $({\rm adj},{\rm adj})$ has a component in both. In any case we will focus on the $O(4)$
structure only and the conclusion at the end will only depend on the $O(4)$ or lattice symmetry
representations.

The $O(4)$ representation in question is the 16 dimensional one 
$1 \,\oplus\, \det \,\oplus\, R \,\oplus\, \det \otimes R \,\oplus\, A$, 
tensored with itself.
These irreducibles, labeled by $(j_1,j_2)$, contain only $j_1 + j_2 \leq 1$. When
forming the tensor square, the irreducible components will then fulfill $j_1 + j_2 \leq 2$. However, the
relevant representations here $H_4, \det\otimes H_4$ and $M_4$ all have $j_1 + j_2 \geq 3$, so they cannot
appear in the space of four-fermion operators, hence no mixing can occur in the continuum.

On the lattice the relevant representations need to be decomposed with respect to $\Wc$ and $\W$. For the
higher moments discussed here, the cubic lattice was not practical even without four-fermion operators so
we will focus on the 16-cell honeycomb only. The $256$ dimensional tensor product representation,
described above, decomposes in each of the two charge conjugation channels as follows,
\bea
C=+1:&& 5\varrho^{(1)}_1\oplus3\varrho^{(1)}_4
\oplus4\varrho^{(4)}_1\oplus4\varrho^{(4)}_4\oplus\varrho^{(4)}_5\oplus3\varrho^{(6)}_1
\oplus\varrho^{(6)}_2\oplus3\varrho^{(9)}_1\oplus\varrho^{(9)}_4\oplus2\varrho^{(16)}_1 \nn \\
C=-1:&& 2\varrho^{(1)}_4\oplus4\varrho^{(4)}_1
\oplus4\varrho^{(4)}_4\oplus6\varrho^{(6)}_1\oplus2\varrho^{(9)}_4
\oplus2\varrho^{(16)}_1.
\eea
The preferred representation for $v_4 = \langle x^3 \rangle$, measurable at rest with continuum-like
renormalization, $\varrho^{(2)}_3$ does not appear in the $C=1$ set, hence does not mix with four-fermion
operators. For $a_3$ we have seen that $\varrho^{(2)}_{2,4}$ renormalize continuum-like with one
non-zero spatial momentum component and these also do not appear in the $C=-1$ set. We also had
$\varrho^{(9)}_4$ and $\varrho^{(12)}_1$ and out of these remaining two $\varrho^{(9)}_4$ does appear,
hence this component mixes with four-fermion operators, while $\varrho^{(12)}_1$ does not. Finally, for
$d_3$ we had three channels and none of them appear above, so none of them mix with four-fermion
operators, in particular the preferred channel available at rest and with continuum-like renormalization,
$\varrho^{(9)}_3$ also does not.

\section{Conclusion and outlook}
\label{conc}

Combined with Wilson coefficients, non-perturbative
lattice results for a set of reduced matrix elements allow the extraction of several properties of
hadrons at low energies. These include the moments of the structure functions
$F_{1,2}(x,Q^2)$ and spin-dependent $g_{1,2}(x,Q^2)$. The operators in question transform in definite
representations of the $O(4)$ continuum Euclidean space-time symmetry group, but any lattice
discretization breaks it. Hence in a lattice calculation the operators are organized according to the
irreducible representations of the finite space-time lattice symmetry group. The larger this symmetry,
the less mixing occurs, which translates directly into computational gain. Ideally, the lattice symmetry
group imposes the same amount of constraints and protection from mixing as the continuum $O(4)$. 
This of course is not possible in general, but {\em is} possible for {\em some} set of operators. In this
work we have shown that the larger symmetry of the 16-cell honeycomb relative to the cubic lattice gives
rise to the ideal continuum-like renormalization pattern for several more operators than was possible on
the cubic lattice. Furthermore, nucleons at rest can be used as well in many of them, which means a much
reduced signal-to-noise problem. 

The analysis involved the flavor non-singlet, flavor singlet, gluonic and four-fermion sectors, 
as well as forward and
off-forward matrix elements. We believe the findings on the qualitative superiority of the 16-cell 
honeycomb over the cubic lattice open the way for dynamical Wilson-fermion simulations to precisions
which were not possible before on the cubic lattice.

In the present work only on-shell matrix elements were considered but the group theoretic classification
was also given for the full set of operators, including those which vanish on-shell. These will be useful
for the study of any off-shell quantity on the lattice in the future such as the actual renormalization
constants in the RI-MOM or SMOM schemes.

\section*{Acknowledgements}

DN would like to thank Will Detmold for suggesting that parton distribution functions
might benefit from the 16-cell honeycomb discretization.
This work was supported by NKFIH grants TKP2021-NKTA-64, 
2026-4.1.1-MISSZIÓK-2026-00005, NKKP Excellence No. 151482 and K-147396.

\end{document}